


\input epsf
\input harvmac.tex

\noblackbox

\def\defn#1{\bigskip\noindent{\bf Definition #1} }

\def\rmk#1{\bigskip\noindent{\bf Remarks} }


\def\figin{\epsfcheck\figin}\def\figins{\epsfcheck\figins}
\def\epsfcheck{\ifx\epsfbox\UnDeFiNeD
\message{(NO epsf.tex, FIGURES WILL BE IGNORED)}
\gdef\figin##1{\vskip2in}\gdef\figins##1{\hskip.5in}
\else\message{(FIGURES WILL BE INCLUDED)}%
\gdef\figin##1{##1}\gdef\figins##1{##1}\fi}
\def\DefWarn#1{}
\def\figinsert{\goodbreak\midinsert}
\def\ifig#1#2#3{\DefWarn#1\xdef#1{fig.~\the\figno}
\writedef{#1\leftbracket fig.\noexpand~\the\figno}%
\figinsert\figin{\centerline{#3}}\medskip\centerline{\vbox{\baselineskip12pt
\advance\hsize by -1truein\noindent\footnotefont{\bf Fig.~\the\figno:} #2}}
\bigskip\endinsert\global\advance\figno by1}


\def\unlockat{\catcode`\@=11}
\def\lockat{\catcode`\@=12}

\unlockat

\def\newsec#1{\global\advance\secno by1\message{(\the\secno. #1)}
\global\subsecno=0\global\subsubsecno=0\eqnres@t\noindent
{\bf\the\secno. #1}
\writetoca{{\secsym} {#1}}\par\nobreak\medskip\nobreak}
\global\newcount\subsecno \global\subsecno=0
\def\subsec#1{\global\advance\subsecno
by1\message{(\secsym\the\subsecno. #1)}
\ifnum\lastpenalty>9000\else\bigbreak\fi\global\subsubsecno=0
\noindent{\it\secsym\the\subsecno. #1}
\writetoca{\string\quad {\secsym\the\subsecno.} {#1}}
\par\nobreak\medskip\nobreak}
\global\newcount\subsubsecno \global\subsubsecno=0
\def\subsubsec#1{\global\advance\subsubsecno by1
\message{(\secsym\the\subsecno.\the\subsubsecno. #1)}
\ifnum\lastpenalty>9000\else\bigbreak\fi
\noindent\quad{\secsym\the\subsecno.\the\subsubsecno.}{#1}
\writetoca{\string\qquad{\secsym\the\subsecno.\the\subsubsecno.}{#1}}
\par\nobreak\medskip\nobreak}

\def\subsubseclab#1{\DefWarn#1\xdef
#1{\noexpand\hyperref{}{subsubsection}%
{\secsym\the\subsecno.\the\subsubsecno}%
{\secsym\the\subsecno.\the\subsubsecno}}%
\writedef{#1\leftbracket#1}\wrlabeL{#1=#1}}
\lockat




\def\IL{\relax{\rm I\kern-.18em L}}
\def\IH{\relax{\rm I\kern-.18em H}}
\def\IR{\relax{\rm I\kern-.18em R}}
\def\IC{\relax\hbox{$\inbar\kern-.3em{\rm C}$}}
\def\IZ{\relax\ifmmode\mathchoice
{\hbox{\cmss Z\kern-.4em Z}}{\hbox{\cmss Z\kern-.4em Z}}
{\lower.9pt\hbox{\cmsss Z\kern-.4em Z}}
{\lower1.2pt\hbox{\cmsss Z\kern-.4em Z}}\else{\cmss Z\kern-.4em Z}\fi}
\def\CA{{\cal A}}
\def\CB {{\cal B}}
\def\CC {{\cal C}}
\def\CM {{\cal M}}
\def\CN {{\cal N}}
\def\CR {{\cal R}}
\def\CD {{\cal D}}
\def\CF {{\cal F}}

\def\CO {{\cal O}}

\def\CZ {{\cal Z}}

\def\CH {{\cal H}}

\font\manual=manfnt \def\dbend{\lower3.5pt\hbox{\manual\char127}}

\def\IZ{\relax\ifmmode\mathchoice
{\hbox{\cmss Z\kern-.4em Z}}{\hbox{\cmss Z\kern-.4em Z}}
{\lower.9pt\hbox{\cmsss Z\kern-.4em Z}}
{\lower1.2pt\hbox{\cmsss Z\kern-.4em Z}}\else{\cmss Z\kern-.4em Z}\fi}
\def\half {{1\over 2}}

\def\p{\partial}
\def\pb{\bar{\partial}}

\def\CM {{\cal M}}
\def\CN {{\cal N}}

\def\CO {{\cal O}}

\def\CK{{\cal K }}

\def\CZ {{\cal Z }}


\def\IZ{\relax\ifmmode\mathchoice
{\hbox{\cmss Z\kern-.4em Z}}{\hbox{\cmss Z\kern-.4em Z}}
{\lower.9pt\hbox{\cmsss Z\kern-.4em Z}}
{\lower1.2pt\hbox{\cmsss Z\kern-.4em Z}}\else{\cmss Z\kern-.4em
Z}\fi}
\def\IB{\relax{\rm I\kern-.18em B}}
\def\IC{{\relax\hbox{$\inbar\kern-.3em{\rm C}$}}}
\def\ID{\relax{\rm I\kern-.18em D}}
\def\IE{\relax{\rm I\kern-.18em E}}
\def\IF{\relax{\rm I\kern-.18em F}}
\def\IG{\relax\hbox{$\inbar\kern-.3em{\rm G}$}}
\def\IGa{\relax\hbox{${\rm I}\kern-.18em\Gamma$}}
\def\IH{\relax{\rm I\kern-.18em H}}
\def\II{\relax{\rm I\kern-.18em I}}
\def\IK{\relax{\rm I\kern-.18em K}}
\def\IP{\relax{\rm I\kern-.18em P}}

\def\IQ{\relax\hbox{$\inbar\kern-.3em{\rm Q}$}}
\def\IP{\relax{\rm I\kern-.18em P}}

\def\IB{\relax{\rm I\kern-.18em B}}
\def\IC{\Bbb{C} }
\def\ID{\relax{\rm I\kern-.18em D}}
\def\IE{\relax{\rm I\kern-.18em E}}
\def\IF{\relax{\rm I\kern-.18em F}}
\def\IG{\relax\hbox{$\inbar\kern-.3em{\rm G}$}}
\def\IGa{\relax\hbox{${\rm I}\kern-.18em\Gamma$}}
\def\IH{\relax{\rm I\kern-.18em H}}
\def\II{\relax{\rm I\kern-.18em I}}
\def\IJ{\relax{\rm I\kern-.18em J}}
\def\IK{\relax{\rm I\kern-.18em K}}
\def\IL{\relax{\rm I\kern-.18em L}}

\def\IN{\relax{\rm I\kern-.18em N}}
\def\IO{\relax{\rm I\kern-.18em O}}
\def\IP{\relax{\rm I\kern-.18em P}}
\def\IQ{\relax\hbox{$\inbar\kern-.3em{\rm Q}$}}
\def\IR{\relax{\rm I\kern-.18em R}}
\def\IW{\relax\hbox{$\inbar\kern-.3em{\rm W}$}}

\def\Im{{\rm Im}}

\def\inbar{\,\vrule height1.5ex width.4pt depth0pt}
\def\mod{\rm mod}
\def\ndt{\noindent}
\def\p{\partial}
\def\pb{{\bar \p}}

\font\cmss=cmss10 \font\cmsss=cmss10 at 7pt
\def\IR{\relax{\rm I\kern-.18em R}}

\def\Tr{\rm Tr}

\def\IC{{\bf C}}


%
\def\inv{^{\raise.15ex\hbox{${\scriptscriptstyle -}$}\kern-.05em 1}}

\def\Dsl{\,\raise.15ex\hbox{/}\mkern-13.5mu D} 
\def\dsl{\raise.15ex\hbox{/}\kern-.57em\partial}

 \def\Tr{{\rm Tr}}




\def\boxit#1{\vbox{\hrule\hbox{\vrule\kern8pt
\vbox{\hbox{\kern8pt}\hbox{\vbox{#1}}\hbox{\kern8pt}}
\kern8pt\vrule}\hrule}}
\def\mathboxit#1{\vbox{\hrule\hbox{\vrule\kern8pt\vbox{\kern8pt
\hbox{$\displaystyle #1$}\kern8pt}\kern8pt\vrule}\hrule}}

\def\windingstrings{Fig. 1}
\def\wilsonline{Fig. 2}
\def\rstates{Fig. 3}


\lref\conway{J.H. Conway and N.J.A. Sloane, {\it Sphere Packings, Lattices,
and Groups}, Springer Verlag, 1993}
%

\lref\AspinwallMN{
P.~S.~Aspinwall,
``K3 surfaces and string duality,''
arXiv:hep-th/9611137.
}
\lref\ClingherUI{
A.~Clingher and J.~W.~Morgan,
``Mathematics underlying the F-theory / heterotic string duality in eight dimensions,''
arXiv:math.ag/0308106.
}

\lref\AharonyTI{
O.~Aharony, S.~S.~Gubser, J.~M.~Maldacena, H.~Ooguri and Y.~Oz,
``Large N field theories, string theory and gravity,''
Phys.\ Rept.\  {\bf 323}, 183 (2000)
[arXiv:hep-th/9905111].
}

\lref\cox{D.A. Cox, {\it Primes of the form $x^2 + n y^2$}, John Wiley, 1989.}

\lref\DijkgraafGF{
R.~Dijkgraaf,
``Instanton strings and hyperKaehler geometry,''
Nucl.\ Phys.\ B {\bf 543}, 545 (1999)
[arXiv:hep-th/9810210].
}
\lref\DijkgraafFQ{
R.~Dijkgraaf, J.~M.~Maldacena, G.~W.~Moore and E.~Verlinde,
``A black hole farey tail,''
arXiv:hep-th/0005003.
}

\lref\DijkgraafXW{
R.~Dijkgraaf, G.~W.~Moore, E.~Verlinde and H.~Verlinde,
``Elliptic genera of symmetric products and second quantized strings,''
Commun.\ Math.\ Phys.\  {\bf 185}, 197 (1997)
[arXiv:hep-th/9608096].
}

\lref\FerraraTW{
S.~Ferrara, G.~W.~Gibbons and R.~Kallosh,
``Black holes and critical points in moduli space,''
Nucl.\ Phys.\ B {\bf 500}, 75 (1997)
[arXiv:hep-th/9702103].
}

\lref\KlebanovME{
I.~R.~Klebanov,
``TASI lectures: Introduction to the AdS/CFT correspondence,''
arXiv:hep-th/0009139.
}

\lref\leshouchesvol{C. Itzykson, J.-M. Luck, P. Moussa, and
M. Waldschmidt, eds.   {\it From Number Theory to Physics},
Springer Verlag, 1995}

\lref\LuUW{
H.~Lu, C.~N.~Pope and E.~Sezgin,
``SU(2) reduction of six-dimensional (1,0) supergravity,''
arXiv:hep-th/0212323.
}

\lref\SeibergXZ{
N.~Seiberg and E.~Witten,
``The D1/D5 system and singular CFT,''
JHEP {\bf 9904}, 017 (1999)
[arXiv:hep-th/9903224].
}


\lref\ManinHN{
Y.~I.~Manin and M.~Marcolli,
``Holography principle and arithmetic of algebraic curves,''
Adv.\ Theor.\ Math.\ Phys.\  {\bf 5}, 617 (2002)
[arXiv:hep-th/0201036].
}

\lref\SkenderisWP{
K.~Skenderis,
``Lecture notes on holographic renormalization,''
Class.\ Quant.\ Grav.\  {\bf 19}, 5849 (2002)
[arXiv:hep-th/0209067].
}

\lref\DenefNB{
F.~Denef,
``Supergravity flows and D-brane stability,''
JHEP {\bf 0008}, 050 (2000)
[arXiv:hep-th/0005049].
}
\lref\DenefXN{
F.~Denef, B.~Greene and M.~Raugas,
``Split attractor flows and the spectrum of BPS D-branes on the quintic,''
JHEP {\bf 0105}, 012 (2001)
[arXiv:hep-th/0101135].
}
\lref\DenefRU{
F.~Denef,
``Quantum quivers and Hall/hole halos,''
JHEP {\bf 0210}, 023 (2002)
[arXiv:hep-th/0206072].
}
\lref\BatesVX{
B.~Bates and F.~Denef,
``Exact solutions for supersymmetric stationary black hole composites,''
arXiv:hep-th/0304094.
}

\lref\LynkerJN{
M.~Lynker, V.~Periwal and R.~Schimmrigk,
``Complex multiplication symmetry of black hole attractors,''
arXiv:hep-th/0303111.
}

\lref\LynkerAJ{
M.~Lynker, V.~Periwal and R.~Schimmrigk,
``Black hole attractor varieties and complex multiplication,''
arXiv:math.ag/0306135.
}
\lref\MillerAG{
S.~D.~Miller and G.~W.~Moore,
``Landau-Siegel zeroes and black hole entropy,''
arXiv:hep-th/9903267.
}

\lref\AshokGK{
S.~Ashok and M.~R.~Douglas,
``Counting Flux Vacua,''
arXiv:hep-th/0307049.
}

\lref\KachruHE{
S.~Kachru, M.~B.~Schulz and S.~Trivedi,
``Moduli stabilization from fluxes in a simple IIB orientifold,''
arXiv:hep-th/0201028.
}

\lref\apostol{T. Apostol, {\it Modular Functions and
Dirichlet Series in Number Theory}, Springer Verlag 1990}
\lref\breckenridge{J.C. Breckenridge,  D.A. Lowe,  R.C. Myers,  A.W. Peet,
A. Strominger, C. Vafa,
``Macroscopic and Microscopic Entropy of Near-Extremal Spinning Black Holes,''
hep-th/9603078; Phys.Lett. B381 (1996) 423-426}
 \lref\carlipi{S. Carlip, ``The $(2+1)$-Dimensional
Black Hole,'' gr-qc/9506079}
\lref\carlipii{S. Carlip and C. Teitelboim}
\lref\cveticyoum{M. Cvetic and D. Youm,
``General Rotating Five Dimensional Black Holes of Toroidally Compactified
Heterotic String,''
hep-th/9603100; Nucl.Phys. B476 (1996) 118-132}
\lref\ez{M. Eichler and D. Zagier, {\it The theory of
Jacobi forms}, Birkh\"auser 1985}
\lref\emss{S. Elitzur, G. Moore,
A. Schwimmer, and N. Seiberg, ``Remarks on the Canonical Quantization of
the Chern-Simons-Witten Theory,''
Nucl. Phys. {\bf B326}(1989)108}
\lref\tkawai{T. Kawai, ``K3 surfaces, Igusa cusp form and
string theory,'' hep-th/9710016}
\lref\kawai{T. Kawai, Y. Yamada, and S. -K. Yang,
``Elliptic genera and $N=2$ superconformal
field theory,'' hepth/9306096 }
\lref\maldacenai{J. Maldacena, ``Black holes and
D-branes,'' hep-th/9705078}
\lref\greybody{J. Maldacena and A. Strominger, Greybody
factors}
\lref\stringexclusion{J. Maldacena and A. Strominger,
``$AdS_3$ black holes and a stringy exclusion
principle,'' hep-th/9804085}
\lref\rademacheri{H. Rademacher, Topics in
Analytic Number Theory}
\lref\rademacherii{H. Rademacher, {\it Lectures on
Elementary Number Theory}, Robert E. Krieger Publishing
Co. , 1964}
\lref\seibschw{N. Seiberg and A. Schwimmer,
``Comments on the N=2,N=3,N=4 superconformal
algebras in two dimensions,'' Phys. Lett.
{\bf 184B}(1987)191 }
\lref\strdec{A. Strominger, December paper}
\lref\townsend{A. Achucarro and P.K. Townsend,
Phys. Lett. {\bf 180B}(1986) 89; J.M. Izquierdo
and P.K. Townsend, ``Supersymmetric spacetimes in
$(2+1)$ adS-supergravity models,'' gr-qc/9501018}
\lref\warner{N. Warner, ``Lectures on $N=2$ superconformal
theories and singularity theory,'' in Superstrings 89}
\lref\wittenads{E. Witten, ``Anti- de Sitter Space and
holography,'' hep-th/9802150; ``Anti-de Sitter space,
thermal phase transition, and confinement in
gauge theories,'' hep-th/9803131}

\lref\vafawitten{C. Vafa and E. Witten, ``A Strong Coupling Test
of S-Duality,'' hep-th/9804074, Nucl.Phys. B431 (1994) 3-77.}


\lref\btz{M.  Ba\~ nados. C.   Teitelboim, and J. Zanelli,
``The Black Hole in Three Dimensional Space Time,''
hep-th/9204099; Phys.Rev.Lett. 69 (1992) 1849-1851}

\lref\bateman{Bateman Manuscript project}

\lref\bost{L. Alvarez-Gaum\'e, J.-B. Bost, G. Moore,
P. Nelson, and C. Vafa, ``Bosonization on higher
genus Riemann surfaces,'' Commun.Math.Phys.112:503,1987 }

\lref\cardi{ K. Behrndt, G. Lopes Cardoso, B. de Wit, R. Kallosh, D. Lvst, T.
Mohaupt, ``Classical and
quantum N=2 supersymmetric black holes,''
hep-th/9610105}

\lref\bloch{S. Bloch, ``The proof of the Mordell
conjecture,'' Math. Intell. {\bf 6}(1984) 41}

\lref\borcherds{R.E. Borcherds, ``Reflection groups of Lorentzian
lattices,'' math.GR/9909123}
 \lref\cassels{J.W.S. Cassels, {\it
Lectures on Elliptic Curves}, Cambridge University Press, 1995}

\lref\cox{D.A. Cox, {\it Primes of the form $x^2 + n y^2$}, John Wiley, 1989.}

\lref\cremona{J.E. Cremona, {\it Algorithms for modular
elliptic curves} Cambridge University Press 1992}
\lref\davenport{H. Davenport, {\it Multiplicative
Number Theory}, Second edition, Springer-Verlag,
GTM 74}

\lref\dewit{Bernard de Wit, Gabriel Lopes Cardoso, Dieter L\"ust,
 Thomas Mohaupt, Soo-Jong Rey,
``Higher-Order Gravitational Couplings and Modular Forms in $N=2,D=4$ Heterotic
String Compactifications,''
hep-th/9607184; G.L. Cardoso, B. de Wit, and
T. Mohaupt, ``Corrections to macroscopic supersymmetric
black hole entropy,'' hep-th/9812082}
\lref\birmingham{D. Birmingham, C. Kennedy,
S. Sen, and A. Wilkins, ``Geometrical finiteness,
holography, and the BTZ black hole,''
hep-th/9812206}

\lref\dvv{R. Dijkgraaf, E. Verlinde, and H. Verlinde,
``Counting Dyons in N=4 String Theory,''
hep-th/9607026;Nucl.Phys. B484 (1997) 543-561}

\lref\duke{W. Duke, ``Hyperbolic distribution problems
and half-integral weight Maass forms,'' Invent. Math.
{\bf 92} (1988) 73}
\lref\faltings{G. Faltings, ``Finiteness theorems for abelian
varieties over number fields,'' in {\it Arithmetic Geometry},
G. Cornell and J.H. Silverman, eds. Springer 1986}
\lref\fks{S. Ferrara, R. Kallosh, and A. Strominger,
``N=2 Extremal Black Holes,''   hep-th/9508072}
\lref\StromingerKF{
A.~Strominger,
``Macroscopic Entropy of $N=2$ Extremal Black Holes,''
Phys.\ Lett.\ B {\bf 383}, 39 (1996)
[arXiv:hep-th/9602111].
}
\lref\fk{S. Ferrara and R. Kallosh, ``Universality of Sypersymmetric
Attractors,''   hep-th/9603090;  ``Supersymmetry and Attractors,''
hep-th/9602136; S. Ferrara, ``Bertotti-Robinson Geometry and Supersymmetry,''
hep-th/9701163}
\lref\enneightbh{ L. Andrianopoli, R. D'Auria,
S. Ferrara, P. Fre', M. Trigiante,
`` $E_{7(7)}$ Duality, BPS Black-Hole Evolution and Fixed Scalars,''
 hep-th/9707087; Nucl.Phys. B509 (1998) 463-518
 }

\lref\segalbourbaki{G. Segal, ``Elliptic cohomology,'' Asterisque {\bf 161-162}(1988)
exp. no. 695, 187-201}
\lref\AlvarezWG{
O.~Alvarez, T.~P.~Killingback, M.~L.~Mangano and P.~Windey,
``String Theory And Loop Space Index Theorems,''
Commun.\ Math.\ Phys.\  {\bf 111}, 1 (1987).
}
\lref\AlvarezDE{
O.~Alvarez, T.~P.~Killingback, M.~L.~Mangano and P.~Windey,
``The Dirac-Ramond Operator In String Theory And Loop Space Index Theorems,''
UCB-PTH-87/11
{\it Invited talk presented at the Irvine Conf. on Non-Perturbative Methods in Physics, Irvine, Calif., Jan 5-9, 1987}
}
\lref\WindeyAR{
P.~Windey,
``The New Loop Space Index Theorems And String Theory,''
 {\it Lectures given at 25th Ettore Majorana Summer School for Subnuclear Physics, Erice, Italy, Aug 6-14, 1987}
}

\lref\fm{S. Ferrara and J. Maldacena, ``Branes, central
charges and $U$-duality invariant BPS conditions,''
hep-th/9706097}

\lref\husemoller{D. Husemoller, {\it Elliptic Curves},
Springer-Verlag, GTM 111}

\lref\faltings{G. Faltings, ``Calculus on arithmetic
surfaces,'' Ann. Math. {\bf 119}(1984) 387}

\lref\bgross{B. Gross, {\it Arithmetic on elliptic
curves with complex multiplication}, Springer-Verlag LNM  776}

\lref\gzsm{B. Gross and D. Zagier,
``On singular moduli,'' J. reine angew. Math.
{\bf 355} (1985) 191}

\lref\kalkol{R. Kallosh and B. Kol,
``E(7) Symmetric Area of the Black Hole Horizon,''
hep-th/9602014}

\lref\knopp{M.I. Knopp, ``Rademacher on $J(\tau)$,
Poincar\'e series of nonpositive weights and the
Eichler cohomology,'' Notices of the Amer. Math.
Soc. {\bf 37}(1990) 385}

\lref\msw{J. Maldacena, A. Strominger, and E. Witten,
``Black Hole Entropy in M-Theory,''  hep-th/9711053}

\lref\arthatt{G. Moore, ``Arithmetic and Attractors,''
hep-th/9807087 ; ``Attractors and Arithmetic,'' hep-th/9807056}
\lref\MoorePN{
G.~W.~Moore,
``Arithmetic and attractors,''
arXiv:hep-th/9807087.
}
\lref\MooreZU{
G.~W.~Moore,
``Attractors and arithmetic,''
arXiv:hep-th/9807056.
}

\lref\WittenHC{
E.~Witten,
``(2+1)-Dimensional Gravity As An Exactly Soluble System,''
Nucl.\ Phys.\ B {\bf 311}, 46 (1988).
}

\lref\lang{S. Lang, {\it Introduction to Arakelov
Geometry}, Springer-Verlag 1988}

\lref\langalg{S. Lang, {\it Algebra}, Addison-Wesley 1970}

\lref\langelliptic{S. Lang, {\it Elliptic Functions},
Addison-Wesley, 1973}

\lref\littlewood{J.E. Littlewood, ``On the class number of
the corpus $P(\sqrt{-k})$,'' Proc. Lond. Math. Soc., Ser. 2, {\bf 27},
Part 5. (1928)}

\lref\mazur{B. Mazur, ``Arithmetic on curves,'' Bull.
Amer. Math. Soc. {\bf 14}(1986) 207}

\lref\miller{S.D. Miller, in preparation.}

\lref\millermoore{S.D. Miller and G. Moore, ``Landau-Siegel Zeroes
and Black Hole Entropy,'' hep-th/9903267}

 \lref\mumford{D. Mumford, Tata Lectures on
Theta}

\lref\peet{A. Peet, ``The Bekenstein Formula and String Theory (N-brane
Theory),''
hep-th/9712253}

\lref\rohrlich{D. Rohrlich, ``Elliptic curves with
good reduction everywhere,'' J. London Math. Soc.,
{\bf 25}(1982)216}

\lref\silvermanag{J. Silverman, {\it Arithmetic Geometry},
G. Cornell and J.H. Silverman, eds. Springer Verlag 1986}

\lref\silveraec{J. Silverman, {\it The Arithmetic of Elliptic
Curves}, Springer Verlag GTM 106 1986}

\lref\silveradvtop{J. Silverman, {\it Advanced Topics in the
Arithmetic of Elliptic Curves} Springer Verlag GTM 151, 1994}

\lref\stromi{A. Strominger, ``Macroscopic Entropy of $N=2$ Extremal Black
Holes,''  hep-th/9602111}

\lref\sv{A. Strominger and C. Vafa,
``Microscopic Origin of the Bekenstein-Hawking Entropy,''
hep-th/9601029; Phys.Lett. B379 (1996) 99-104 }

\lref\tatuzawa{T. Tatuzawa, ``On a theorem of
Siegel,'' Jap. J. Math. {\bf 21}(1951)163 }

\lref\weil{A. Weil, {\it Elliptic functions according to
Eisenstein and Kronecker}, Springer-Verlag 1976}

\lref\wittenls{E. Witten, `` Phases of $N=2$ Theories In Two Dimensions,''
hep-th/9301042;Nucl.Phys. B403 (1993) 159-222}

\lref\review{  O. Aharony, S.S. Gubser, J. Maldacena, H. Ooguri, Y. Oz,
``Large N Field Theories, String Theory and Gravity,'' hep-th/9905111. }
\lref\cvl{M. Cveti\'c and F. Larsen, ``Near Horizon Geometry
of Rotating Black Holes in Five Dimensions,''
hep-th/9805097}

\lref\martsahak{E. Martinec and V. Sahakian,
``Black holes and five-brane thermodynamics,''
hep-th/9901135; Phys. Rev. {\bf D60}(1999)
064002}
\lref\carlipsum{S. Carlip,
``The sum over topologies in three-dimensional
Euclidean quantum gravity,''
hep-th/9206103; Class.Quant.Grav.10:207-218,1993.
S. Carlip, ``Dominant topologies in
Euclidean quantum gravity,''
gr-qc/9710114; Class.Quant.Grav.15:2629-2638,1998.
 }
\lref\degersezgin{S. Deger, A. Kaya, E. Sezgin, and P. Sundell,
``Spectrum of D=6, N=4b supergravity on $AdS_3 \times S^3$,''
hep-th/9804166}
 \lref\hardyram{G.H. Hardy and S. Ramanujan, ``Asymptotic
formulae in combinatory analysis,'' Proc. Lond. Math. Soc. {\bf
2}(1918)75} \lref\tftads{E. Witten, ``AdS/CFT correspondence and
Topological field theory,'' hep-th/9812012} \lref\jones{E.~
Witten, ``Quantum Field Theory and  the Jones Polynomial", Comm.
Math. Phys. {\bf 121} (1989) 351.} \lref\deboer{J.
de Boer,``Six-dimensional supergravity on $S^3 \times AdS_3$ and 2d
conformal field theory,'' hep-th/9806104; ``Large N Elliptic Genus
and AdS/CFT Correspondence,'' hep-th/9812240}

\lref\carlipbook{S. Carlip, {\it Quantum gravity in 2+1
dimensions}, Cambridge University Press, 1998}
\lref\eguchitaormina{T. Eguchi and A. Taormina, ``Character
formulas for the N=4 superconformal algebra,'' Phys. Lett. {\bf
200B}(1988) 315.}
 \lref\elstrodt{J. Elstrodt, F. Grunewald,
and J. Mennicke, {\it Groups acting on hyperbolic space}, Springer
1998}
 \lref\gottschsoergel{L.
G\"ottsche and W.  Soergel, ``Perverse sheaves and the cohomology
of Hilbert schemes of smooth algebraic surfaces,'' Math. Ann. {\bf
296} (1993)235 }
\lref\hennschwimm{M. Henneaux, L. Maoz, and A. Schwimmer,
``Asymptotic dynamics and asymptotic symmetries of
three-dimensional extended AdS supergravity,'' hep-th/9910013}.

\lref\littlewood{J.E. Littlewood, {\it Littlewood's Miscellany},
Cambridge Univ. Press, 1986} \lref\gritsenko{V. Gritsenko,
``Complex vector bundles and Jacobi forms,'' math.AG/9906191;
``Elliptic genus of Calabi-Yau manifolds and Jacobi and Siegel
modular forms,'' math.AG/9906190}
\lref\borisov{L.A. Borisov and A, Libgober, ``Elliptic Genera
and Applications to Mirror Symmetry,'' math.AG/9904126}

\lref\dmvv{R. Dijkgraaf, G.
Moore, E. Verlinde and H. Verlinde, ``Elliptic Genera of Symmetric
Products and Second Quantized Strings,'' Commun.Math.Phys. 185
(1997) 197-209} \lref\mms{J. Maldacena, G. Moore, and A.
Strominger, ``Counting BPS Blackholes in Toroidal Type II String
Theory,'' hep-th/9903163 }
\lref\zoo{G. Moore and N. Seiberg, ``Taming the conformal zoo,''
Phys. Lett. {\bf 220B} (1989) 422}
 \lref\Iwaniec{H. Iwaniec, {\it
Topics in Classical Automorphic Forms}, AMS Graduate Studies in
Math. {\bf 17} 1997; {\it Introduction to the Spectral Theory of
Automorphic Forms}, Revista Mathematica Iberoamericana, 1995}
 \lref\townsend{A. Achucarro and P.K.
Townsend, ``A Chern-Simons Action for Three-Dimensional
Anti-de-Sitter Theories,'' Phys. Lett. {\bf B180}(1986)89}
 \lref\maldasusskind{Juan M. Maldacena
and Leonard Susskind, "D-branes and fat black
holes",hep-th/9604042, Nucl. Phys {\bf B475}(1996), 679-690}
\lref\petersson{H. Petersson, ``\"Uber die Entwicklungskoeffizienten
der automorphen Formen,'' Acta. Math. {\bf 58}(1932) 169}
\lref\radpaper{H. Rademacher, ``The Fourier coefficients of the
modular invariant $J(\tau)$,'' Amer. J. Math. {\bf 60}(1938)501}

\lref\sarnak{P. Sarnak, {\it Some applications of modular forms},
Cambridge 1990. }

 \lref\witteneg{E. Witten, ``Elliptic Genera and
Quantum Field Theory,'' Commun. Math. Phys. {\bf 109}(1987)525;
``The index of the Dirac operator in loop space,'' Proceedings of
the conference on elliptic curves and modular forms in algebraic
topology, Princeton NJ, 1986.}

\lref\connes{A. Connes, {\it Noncommutative Geometry},
Academic Press (1994).}
%

\lref\KalloshUY{
R.~Kallosh and B.~Kol,
``E(7) Symmetric Area of the Black Hole Horizon,''
Phys.\ Rev.\ D {\bf 53}, 5344 (1996)
[arXiv:hep-th/9602014].
}

\lref\HullYS{
C.~M.~Hull and P.~K.~Townsend,
``Unity of superstring dualities,''
Nucl.\ Phys.\ B {\bf 438}, 109 (1995)
[arXiv:hep-th/9410167].
}

\lref\ArutyunovBY{
G.~Arutyunov, A.~Pankiewicz and S.~Theisen,
``Cubic couplings in D = 6 N = 4b supergravity on AdS(3) x S(3),''
Phys.\ Rev.\ D {\bf 63}, 044024 (2001)
[arXiv:hep-th/0007061].
}

\lref\shioda{T. Shioda and H. Inose, ``On singular
K3 surfaces,'' in Complex analysis and algebraic geometry,
Cambridge University Press, Cambridge, 1977}

\lref\wendlandthesis{K. Wendland, ``Moduli spaces of unitary
conformal field theories,'' Ph. D. Thesis, University of Bonn,
BONN-IR-2000-11}

\lref\WendlandMA{
K.~Wendland,
``On Superconformal Field Theories Associated to Very Attractive Quartics,''
arXiv:hep-th/0307066.
}
\lref\HosonoYB{
S.~Hosono, B.~H.~Lian, K.~Oguiso and S.~T.~Yau,
``Classification of c = 2 rational conformal field theories via the Gauss  product,''
arXiv:hep-th/0211230.
}
\lref\GukovNW{
S.~Gukov and C.~Vafa,
``Rational conformal field theories and complex multiplication,''
arXiv:hep-th/0203213.
}

\lref\TripathyQW{
P.~K.~Tripathy and S.~P.~Trivedi,
``Compactification with flux on K3 and tori,''
JHEP {\bf 0303}, 028 (2003)
[arXiv:hep-th/0301139].
}

\lref\yui{N. Yui,
`Update on the modularity of Calabi-Yau varieties,''
Fields Communication Series Vol. 38 (2003), pp. 307-362.
American Math. Soc.}

\lref\LarsenUK{
F.~Larsen and E.~J.~Martinec,
JHEP {\bf 9906}, 019 (1999)
[arXiv:hep-th/9905064].
}

\lref\LarsenDH{
F.~Larsen and E.~J.~Martinec,
``Currents and moduli in the (4,0) theory,''
JHEP {\bf 9911}, 002 (1999)
[arXiv:hep-th/9909088].
}

\lref\BanksIA{
T.~Banks,
``Supersymmetry, the cosmological constant and a theory of quantum  gravity in our universe,''
arXiv:hep-th/0305206.
}
\lref\BanksVP{
T.~Banks,
``A critique of pure string theory: Heterodox opinions of diverse  dimensions,''
arXiv:hep-th/0306074.
}

\lref\HorowitzXK{
G.~T.~Horowitz and D.~Marolf,
``A new approach to string cosmology,''
JHEP {\bf 9807}, 014 (1998)
[arXiv:hep-th/9805207].
}

\lref\MartinecCF{
E.~J.~Martinec and W.~McElgin,
``String theory on AdS orbifolds,''
JHEP {\bf 0204}, 029 (2002)
[arXiv:hep-th/0106171].
}
\lref\MartinecXQ{
E.~J.~Martinec and W.~McElgin,
``Exciting AdS orbifolds,''
JHEP {\bf 0210}, 050 (2002)
[arXiv:hep-th/0206175].
}

\lref\LiuFT{
H.~Liu, G.~Moore and N.~Seiberg,
``Strings in a time-dependent orbifold,''
JHEP {\bf 0206}, 045 (2002)
[arXiv:hep-th/0204168].
}
\lref\LiuKB{
H.~Liu, G.~Moore and N.~Seiberg,
``Strings in time-dependent orbifolds,''
JHEP {\bf 0210}, 031 (2002)
[arXiv:hep-th/0206182].
}
\lref\LiuYD{
H.~Liu, G.~Moore and N.~Seiberg,
``The challenging cosmic singularity,''
arXiv:gr-qc/0301001.
}
\lref\HorowitzMW{
G.~T.~Horowitz and J.~Polchinski,
``Instability of spacelike and null orbifold singularities,''
Phys.\ Rev.\ D {\bf 66}, 103512 (2002)
[arXiv:hep-th/0206228].
}

\lref\KrausIV{
P.~Kraus, H.~Ooguri and S.~Shenker,
``Inside the horizon with AdS/CFT,''
Phys.\ Rev.\ D {\bf 67}, 124022 (2003)
[arXiv:hep-th/0212277].
}

\lref\VafaXN{
C.~Vafa,
``Evidence for F-Theory,''
Nucl.\ Phys.\ B {\bf 469}, 403 (1996)
[arXiv:hep-th/9602022].
}

\lref\petridis{Y. Petridis and M. Skarsholm Risager,
``Modular symbols have a normal distribution,'' preprint.}

\lref\CandelasDM{
P.~Candelas, X.~De La Ossa, A.~Font, S.~Katz and D.~R.~Morrison,
``Mirror symmetry for two parameter models. I,''
Nucl.\ Phys.\ B {\bf 416}, 481 (1994)
[arXiv:hep-th/9308083].
}

\lref\FriedmanYQ{
R.~Friedman, J.~Morgan and E.~Witten,
``Vector bundles and F theory,''
Commun.\ Math.\ Phys.\  {\bf 187}, 679 (1997)
[arXiv:hep-th/9701162].
}
\lref\FriedmanSI{
R.~Friedman, J.~W.~Morgan and E.~Witten,
``Principal G-bundles over elliptic curves,''
Math.\ Res.\ Lett.\  {\bf 5}, 97 (1998)
[arXiv:alg-geom/9707004].
}

\lref\GukovYA{
S.~Gukov, C.~Vafa and E.~Witten,
``CFT's from Calabi-Yau four-folds,''
Nucl.\ Phys.\ B {\bf 584}, 69 (2000)
[Erratum-ibid.\ B {\bf 608}, 477 (2001)]
[arXiv:hep-th/9906070].
}

\lref\CurioAE{
G.~Curio, A.~Klemm, B.~Kors and D.~Lust,
``Fluxes in heterotic and type II string compactifications,''
Nucl.\ Phys.\ B {\bf 620}, 237 (2002)
[arXiv:hep-th/0106155].
}
\lref\CurioSC{
G.~Curio, A.~Klemm, D.~Lust and S.~Theisen,
``On the vacuum structure of type II string compactifications on  Calabi-Yau spaces with H-fluxes,''
Nucl.\ Phys.\ B {\bf 609}, 3 (2001)
[arXiv:hep-th/0012213].
}

\lref\mmda{``Melvin models and Diophantine Approximation,'' to appear}

\lref\emillectures{E. Martinec,
Lectures given at the Komaba workshop, November 1999;
notes available at
http://theory.uchicago.edu/$\sim$ejm/japan99.ps}

\lref\MaldacenaRF{
J.~Maldacena and L.~Maoz,
``Wormholes in AdS,''
arXiv:hep-th/0401024.
}

\lref\LuninIZ{
O.~Lunin, J.~Maldacena and L.~Maoz,
``Gravity solutions for the D1-D5 system with angular momentum,''
arXiv:hep-th/0212210.
}

\lref\ClingherUI{
A.~Clingher and J.~W.~Morgan,
``Mathematics underlying the F-theory / heterotic string duality in eight
dimensions,''
arXiv:math.ag/0308106.
}

\lref\GiryavetsVD{
A.~Giryavets, S.~Kachru, P.~K.~Tripathy and S.~P.~Trivedi,
``Flux compactifications on Calabi-Yau threefolds,''
arXiv:hep-th/0312104.
}

\lref\BalasubramanianKQ{
V.~Balasubramanian, A.~Naqvi and J.~Simon,
``A multi-boundary AdS orbifold and DLCQ holography: A universal holographic
description of extremal black hole horizons,''
arXiv:hep-th/0311237.
}

\lref\LynkerHJ{
M.~Lynker, R.~Schimmrigk and S.~Stewart,
``Complex Multiplication of Exactly Solvable Calabi-Yau Varieties,''
arXiv:hep-th/0312319.
}

\lref\LianZV{
B.~H.~Lian and S.~T.~Yau,
``Arithmetic properties of mirror map and quantum coupling,''
Commun.\ Math.\ Phys.\  {\bf 176}, 163 (1996)
[arXiv:hep-th/9411234].
}

\lref\SenBP{
A.~Sen,
``Orientifold limit of F-theory vacua,''
Nucl.\ Phys.\ Proc.\ Suppl.\  {\bf 68}, 92 (1998)
[Nucl.\ Phys.\ Proc.\ Suppl.\  {\bf 67}, 81 (1998)]
[arXiv:hep-th/9709159].
}

\lref\douglasdenef{F. Denef and M. Douglas, to appear}

\lref\LercheCA{
W.~Lerche and N.~P.~Warner,
``Index Theorems In N=2 Superconformal Theories,''
Phys.\ Lett.\ B {\bf 205}, 471 (1988).
}

\lref\LercheQK{
W.~Lerche, B.~E.~W.~Nilsson, A.~N.~Schellekens and N.~P.~Warner,
``Anomaly Cancelling Terms From The Elliptic Genus,''
Nucl.\ Phys.\ B {\bf 299}, 91 (1988).
}

\lref\PilchGS{
K.~Pilch and N.~P.~Warner,
``String Structures And The Index Of The Dirac-Ramond Operator On Orbifolds,''
Commun.\ Math.\ Phys.\  {\bf 115}, 191 (1988).
}

\lref\PilchEN{
K.~Pilch, A.~N.~Schellekens and N.~P.~Warner,
``Path Integral Calculation Of String Anomalies,''
Nucl.\ Phys.\ B {\bf 287}, 362 (1987).
}

\lref\SchellekensYJ{
A.~N.~Schellekens and N.~P.~Warner,
``Anomaly Cancellation And Selfdual Lattices,''
Phys.\ Lett.\ B {\bf 181}, 339 (1986).
}

\lref\SchellekensYI{
A.~N.~Schellekens and N.~P.~Warner,
``Anomalies And Modular Invariance In String Theory,''
Phys.\ Lett.\ B {\bf 177}, 317 (1986).
}

\lref\KawaiTE{
  T.~Kawai,
  ``String duality and modular forms,''
  Phys.\ Lett.\  B {\bf 397}, 51 (1997)
  [arXiv:hep-th/9607078].
}

\lref\ManschotHA{
  J.~Manschot and G.~W.~Moore,
  ``A Modern Farey Tail,''
  arXiv:0712.0573 [hep-th].
}


\rightline{RUNHETC-2003-36 }
\Title{
\rightline{hep-th/0401049}}
{\vbox{\centerline{Les Houches Lectures on Strings and Arithmetic$^*$}}}
\bigskip
\centerline{  Gregory W. Moore\footnote{}{$^*$ Summary of
lectures delivered at the conference {\it Number Theory,
Physics,  and Geometry } Les Houches, March, 2003}}

\bigskip
\centerline{ { Department of Physics, Rutgers University}}
\centerline{ Piscataway, NJ 08854-8019, USA}

\bigskip
\noindent
These are lecturenotes for two lectures delivered at the
Les Houches workshop on Number Theory, Physics, and Geometry,
March 2003.  They review  two examples of
interesting interactions between number theory
and string compactification, and raise some new questions and
issues in the context of those examples. The first example
concerns the role of the Rademacher expansion of
coefficients of modular forms in the AdS/CFT correspondence.
The second example concerns the role of the
``attractor mechanism'' of supergravity in selecting
certain arithmetic Calabi-Yau's as distinguished compactifications.

\vfill

\Date{Jan. 6, 2004 }

\newsec{Introduction}

Several of the most interesting developments of
modern string theory use some of the mathematical
tools of modern number theory. One striking example
of this is  the importance of arithmetic groups in
the theory of duality symmetries. Another
example, somewhat related, is the occurance of
automorphic forms for arithmetic groups in low energy
effective supergravities.  These examples are quite well-known.

In the following two lectures we explore two other
less-well-known examples of curious roles
of number theoretic techniques in   string theory.
The first concerns a technique of analytic number theory
and its role in the AdS/CFT
correspondence. The second is related to the ``attractor
equations.'' These are equations
on Hodge structures of Calabi-Yau manifolds and have arisen
in a number of contexts connected with string compactification.
Another topic of possible interest to readers of this volume
will appear elsewhere \mmda.

\newsec{Potential Applications of the AdS/CFT Correspondence
to Arithmetic}

\subsec{Summary}

In this talk we are going to indicate how the ``AdS/CFT correspondence''
of string theory might have some interesting relations to analytic
number theory. The main part of the talk reviews work done with
R. Dijkgraaf, J. Maldacena, and E. Verlinde which appeared in
\DijkgraafFQ. Ideas similar in spirit, but, so far as I know, different in
detail have appeared in \ManinHN.

\subsec{Summary of the AdS/CFT correspondence}

The standard reviews on the AdS/CFT correspondence are \refs{\AharonyTI ,\KlebanovME, \SkenderisWP}.
In this literature,
``anti-deSitter space''  comes in two signatures. The
Euclidean version is simply hyperbolic space:
\eqn\ah{
AdS_{n+1} = \IH^{n+1} = SO(1,n+1)/SO(n+1)
}
while the Lorentzian version is
\eqn\ahi{
AdS_{1,n} = SO(2,n)/SO(1,n)
}
where on the right-hand side we should take the   universal cover.
These spacetimes are nice solutions to Einstein's equations with negative cosmological constant.
\eqn\ahii{
\CR_{\mu\nu} - \half g_{\mu\nu} \CR + \Lambda g_{\mu\nu} =0 \qquad  \Lambda = -1
}
In the context of string theory they arise very naturally in certain solutions
to 10- and 11-dimensional supergravity associated with configurations of branes.

Some  important examples (by no means all) of such solutions include
\item{1.} $AdS_2 \times S^2 \times M_6$
where $M_6$ is a Calabi-Yau $3$-fold.  The associated D-brane configurations
are discussed in Lecture II below.

\item{2.} $AdS_3 \times S^3 \times M_4$ where $M_4$ is   a $K3$ surface or a
torus $T^4$,  or $S^3\times S^1$.

\item{3.} $AdS_5 \times S^5 $. This is the geometry
associated to a large collection of coincident $D3$ branes in
10-dimensional Minkowski space and is the subject of much of
the research done in AdS/CFT duality.

 At the level of slogans the AdS/CFT  conjecture states that
{\it 10-dimensional string theory on
\eqn\a{AdS_{n+1}\times K }
is ``equivalent'' to
a super-conformal field theory -- i.e., a QFT without gravity -- on the conformal boundary
\eqn\aa{\p AdS_{n+1} .  }
}
The   ``conformal boundary of AdS'' means, operationally,
\eqn\aaaa{
\p AdS_{n+1} = S^n \qquad or \qquad S^{n-1} \times \IR
}
More fundamentally it is the conformal boundary in the  sense of Penrose.

Of course, the above slogan is extremely vague.
One goal of this talk is to give an example where the statement can be
made mathematically quite precise. We are  explaining this example in the
present volume because it   involves some interesting analytic
number theory. The  hope is that a precise version of the AdS/CFT principle
can eventually be turned into a useful tool in number theory, and
the present example is adduced as evidence for this hope. At the end
of the talk we will make some more speculative suggestions along these lines.

\subsubsec{AdS/CFT made a little more precise}

In order to explain our example it is necessary to
 make the statement of AdS/CFT a little more precise.

Consider 10D string theory on
$X$ which is a noncompact manifold which at infinity looks locally like
\eqn\a{
X \sim AdS_{n+1} \times K
}
Let us think  of string theory as an infinite-component field theory on this spacetime.
In particular the fields include the graviton  $g_{\mu\nu}$, as well as
(infinitely) many others. Let us denote the generic field by $\phi$.
We assume there is a well-defined notion of a partition function of
string theory associated to this background. Schematically, it should be
some kind of functional integral:
\eqn\funint{
Z_{string} = \int [dg_{\mu\nu}] [d \phi] \cdots e^{-\int \sqrt{g} \CR(g) + (\nabla \phi)^2 + \cdots }
}
Even at this schematic level we can see one crucial   aspect of the functional integral:
we must specify the boundary conditions of the fields at infinity.

Since spacetime has a factor which is locally $AdS$ at infinity there is a
second order pole in the metric at infinity.
Let $r$ denote a coordinate so that the conformal boundary is at $r \to \infty$
and such that the metric takes the asymptotic form
\eqn\a{
ds^2_X  \rightarrow {dr^2 \over r^2} + r^2 \hat g_{ij}(\theta) d\theta^i d \theta^j + ds^2_K
}
where $\theta^i$ denote coordinates on $S^n$. In these coordinates we impose
boundary conditions on the remaining fields:
\eqn\a{
\phi(r,\theta) \rightarrow r^h \phi_0(\theta)
}

The functional integral \funint\ is thus   a function
\foot{In fact, it should be considered as a ``wavefunction.''  In the closely
related Chern-Simons gauge theory/RCFT duality this is literally true. }    of the boundary data:
\eqn\vwv{
Z_{string}(\hat g,\phi_0, \dots)
}

We can now state  slightly  more precise versions of AdS/CFT. There is a
slightly different formulation for Euclidean and Lorentzian signature.

The Euclidean version of AdS/CFT states that there   exists a CFT $\CC$ defined on $\p AdS_{n+1} = S^n $
such that the space $\CA$ of local operators in $\CC$ is dual to the
string theory boundary conditions:
\eqn\dualop{
\phi_0 \rightarrow  \Phi_{\phi_0} \in \CA
}
such that
\eqn\a{
\left\langle e^{\int_{S^n}   \Phi_{\phi_0}(\theta) } \right\rangle_{CFT }
= Z_{string}(\hat g, \phi_0,\dots)
}
This statement of the AdS/CFT correspondence, while conceptually simple, is
quite oversimplified. Both sides of the equation are infinite, must be regularized, etc.
See the above cited reviews for a somewhat more careful discussion.

The  Lorentzian version of AdS/CFT states that there
is an isomorphism of Hilbert spaces between the gravity and
CFT formulations that preserves certain operator algebras.
These are $\CH_{\CC}$, the Hilbert space of the CFT $\CC$ on $S^{n-1}\times \IR$,
and  $\CH_{string}$, the Hilbert space of string (or M) theory on $AdS_{n+1} \times K$.
This is already a nontrivial statement when one considers both
sides as representations of the superconformal group.
  An approximation to $\CH_{string}$ is given by particles in the supergravity
approximation, and corresponding states in the CFT have been found.
See \refs{\AharonyTI}. Whether or not the isomorphism truly holds for the
entire Hilbert space is problematic because of multi-particle states and
because of the role of black holes. Indeed, it is clear that one {\it must}
include quantum states in $\CH_{string}$ associated both to black holes
and to strings and D-branes in order to avoid contradictions.

\subsec{A particular example}

In the remainder of this talk we will focus on
the example of type  $IIB$ string theory on $AdS_3 \times S^3 \times K3$.
In this case the   dual CFT on $\p AdS_3$ is a two-dimensional CFT $\CC$.

{}From symmetry considerations it is clear that the dual CFT has
 $(4,4)$ supersymmetry. It is thought that $\CC$   admits marginal deformations
to a supersymmetric $\sigma$-model whose  target
space $X$   is a hyperkahler resolution
\eqn\hkres{
X  \rightarrow (K3)^k/S_k = {\rm Sym}^k(K3).
}
In comparing the gravity and CFT side we make the
identification
\eqn\a{
k = \ell/4G
}
where
$\ell$ is the  radius of $S^3$ (which in turn is the
curvature radius of $AdS_3$), while
$G$ is the Newton constant in 3 dimensions. The quantization
of $\ell/4G$ can be seen intrinsically on the gravity
side from the existence of certain Chern-Simons couplings
for $SU(2)$ gauge fields with coefficient $k$.

The ``proof'' of the correspondence proceeds by
 studying the near horizon geometry of solutions of the
supergravity equations representing $Q_1$ D1 branes
and $Q_5$ D5 branes wrapping $K3 \times S^1$. One
studies the low energy excitations of the ``string'' wrapping
the $S^1$ factor. The   dynamics of these excitations
are   are described by a supersymmetric nonlinear sigma model with
target space \hkres\  for  $k=Q_1 Q_5 + 1$. The moduli space of
supergravity solutions, as well as the moduli space of the
supersymmetric sigma model are both the space
\eqn\mdspc{
\Gamma \backslash SO(4,21)/SO(4) \times SO(21)
}
where $\Gamma$ is an arithmetic subgroup of $SO(4,21;\IZ)$.
See  \refs{\DijkgraafGF,\SeibergXZ,\LarsenUK,\LarsenDH,\emillectures} for some
explanation of the details of this.

The correlation function whose equivalence in AdS and CFT
formulations we wish to present is  a certain parititon
function which, on the CFT side is the
 {\it elliptic genus} of the conformal field theory.
The reason we focus on this quantity is that the
dual CFT is very subtle.
The elliptic genus
 is a ``correlation  function'' of the CFT $\CC$ which is invariant
under many perturbations of the CFT, and is therefore robust and
computable.  Nevertheless, the resulting function is also still
 nontrivial and contains much
useful information.

Our strategy will be to write the elliptic genus in a form
that makes the connection
to quantum gravity on $AdS_3$ clear.
 The form in which we can make this connection is a Poincar\'e series
for the elliptic genus.

\subsec{ Review of Elliptic Genera}

For some background on the elliptic genus, see
 \refs{\SchellekensYI,\SchellekensYJ,\PilchEN,\witteneg,
\AlvarezWG,\AlvarezDE,\WindeyAR,\PilchGS,\LercheQK,\LercheCA,\KawaiJK}.

Let $\CC$ be a  CFT with $(2,2)$ supersymmetry. This means the
 Hilbert space $\CH$ is a representation
of superconformal $Vir^{\CN=2}_{left} \oplus Vir^{\CN=2}_{right}$,
where the subscript refers to the usual separation of conformal fields
into left- and right-moving components.

Let us recall that the $\CN=2$ superconformal algebra is generated
by Virasoro operators $L_n$, and $U(1)$ current algebra $ J_n$, with
 $n \in \IZ$, and superconformal generators $G^\pm_r $ with $r\in \IZ+\half $
for the NS algebra and $r\in \IZ $ for the R algebra. The
commutation relations are:
\eqn\vir{
[L_n, L_m ] = (n-m) L_{n+m} + {c\over 12} (n^3-n) \delta_{n+m,0}
}
\eqn\gee{
[G_r^\pm, G_s^\mp] = 2 L_{r+s} + (r-s) J_{r+s} + {c\over 12}(4r^2-1) \delta_{r+s,0}
}
\eqn\jay{
[J_n,J_m] = {c\over 3}n \delta_{n+m,0}
}
\eqn\lgee{
[L_n, G_r^\pm]  = (\half n-r )G^\pm_{n+r}
}
\eqn\a{
[J_n, G_r^\pm ] = \pm G_{n+r}^\pm
}
\eqn\a{
[L_n, J_m] = -m J_{n+m}
}
Right-moving generators are denoted $ \tilde L_n, \tilde J_n, \tilde G_r^\pm$.

The elliptic genus is
\eqn\ellgen{
\chi( \tau,z) :=
{\Tr}_{RR} e^{2 \pi i \tau  (L_0-c/24) }
 e^{2 \pi i \tilde \tau (\tilde  L_0-c/24) }
 e^{2 \pi i z  J_0}    (-1)^{F }
}
where the trace is in the Ramond-Ramond sector and $(-1)^F = e^{i \pi (J_0 - \tilde J_0)}$.

In a unitary $(2,2)$ superconformal field theory the operators
$L_0, \tilde L_0, J_0, \tilde J_0$ may be simultaneously diagonalized.
In a unitary theory the spectrum satisfies $L_0 - c/24 \geq 0 $ in the
Ramond sector (and similarly for the right-movers). States with
$\tilde L_0 = c/24$ are called {\it right-BPS}. It follows straightforwardly
from the commutation relations \gee\ that only right-BPS states
  make a nonzero
contribution to the trace \ellgen\  and hence $\chi(\tau,z)$ has
Fourier expansion
\eqn\expcffs{
\sum_{n\geq 0, r} c(n,r) q^n y^r
}
where $q= e^{2\pi i \tau}$ and $  y = e^{2\pi i z}$.

In this paper we will be considering superconformal theories
with $(4,4)$ supersymmetry. These are special cases of the $(2,2)$ theories,
but have extra structure: For each chirality, left and right,
the  $U(1)$ current algebra \jay\  is
enhanced to a level $k$ affine $SU(2)$ current algebra $T_n^a$.
In addition, for each chirality, there is a global $SU(2)$ symmetry $\hat T^a$  and the
four  supercharges transform in the $(\half,\half)$ representation
of the global $SU(2) \times SU(2)$.  The
Virasoro central charge is given by $c=6k$.

\subsubsec{Properties of the Elliptic Genus}

The elliptic genus satisfies two key properties:
modular invariance and spectral flow invariance.
The modular invariance follows from the fact that
$\chi(\tau,z)$ can be regarded as a path integral of $\CC$ on
a two-dimensional torus $S^1  \times S^1$
with odd spin structure for the fermions.

%
%
%

Under modular transformations
\eqn\modinv{
\chi({a \tau + b \over  c \tau + d} , {z \over  c \tau + d})
=  e^{ 2 \pi i k { c z^2 \over  c \tau + d} } \chi(\tau,z)
}
In order to prove this from the path integral viewpoint
 note that including the parameter
$z$ involves adding a term $\sim \int \bar A \wedge J$ to the
worldsheet action. From the singular ope of $J$ with itself
one needs to include a subtraction term.  After making a
modular transformation this subtraction term must change, the
difference is finite and accounts for the exponential prefactor in
\modinv.

The $\CN=2$ algebra has a well-known   spectral flow isomorphism \seibschw
\eqn\a{
\eqalign{
  G_{n \pm a}^\pm   & \rightarrow
G^\pm_{n \pm (a + \theta)} \cr
L_0 &
 \rightarrow L_0 + \theta J_0 + \theta^2 k \cr
J_0 & \rightarrow
J_0 + 2 \theta k \cr}
}
which
implies that
\eqn\specflow{
\chi(\tau,z+ \ell \tau + m) = e^{-2 \pi i k (\ell^2 \tau+ 2 \ell z)}
\chi(\tau,z)  \qquad \ell,m \in \IZ
}

The identities \modinv\   and \specflow\  above are summarized in the mathematical
definition \ez:

\defn{}
 A {\it weak Jacobi form} $\phi(\tau,z)$ of weight $w$ and index $k$
satisfies the identities: \eqn\a{ \phi({a \tau + b \over  c \tau +
d} , {z \over  c \tau + d}) = (c \tau+d)^w e^{ 2 \pi i k { c z^2
\over  c \tau + d} } \phi(\tau,z) }
\eqn\ab{ \phi(\tau,z+ \ell \tau
+ m) = e^{-2 \pi i k (\ell^2 \tau+ 2 \ell z)} \phi(\tau,z)  \qquad
\ell,m\in \IZ } and has a Fourier expansion with $c(n,r)=0$ unless
$n\geq 0$.

Thus, the elliptic genus of a unitary $(4,4)$ superconformal
field theory is a weak Jacobi form of weight $0$ and level $k$.
Much useful information on Jacobi forms can be found in \ez.

Using the spectral flow identity \ab\ we find that $c(n,\ell) = c(n+
\ell s + k s^2,  \ell + 2k s) $, for $s$ an integer, and therefore
$c(n,\ell) = c(n-\nu s_0 - k s_0^2, \nu):=c_\nu(4kn-\ell^2)$ if
$\ell = \nu + 2k s_0$. Using this it is straightforward to derive
 \eqn\paradec{
\chi(\tau,z) =  \sum_{\mu=-k +1}^{k } h_\mu(\tau)
\Theta_{\mu,k}(z,\tau) } Here $\Theta_{\mu,k}(z,\tau)$ are level $k$
theta functions \eqn\thetmk{ \eqalign{ \Theta_{\mu,k}(z,\tau) & :=
\sum_{\ell\in \IZ, \ell = \mu \mod 2k } q^{\ell^2/(4k)} y^{\ell} \cr
& = \sum_{n\in \IZ} q^{k(n+\mu/(2k))^2} y^{ (\mu + 2k n)} \cr} }
 We denote the combinations even and odd in $z$ by $\Theta^\pm_{\mu,k}$.

Our goal now is to write the elliptic genus for the
conformal field theory appearing in the AdS/CFT correspondence
in a fashion suitable for interpretation via AdS/CFT.  This fashion
will simply be a Poincar\'e series. Before doing this in section 2.5
we make a small digression.

 \subsubsec{  Digression 1: Elliptic Genera for Symmetric Products }

If the conformal field theory $\CC$ is a sigma model
with target space $X$,  denoted $\CC = \sigma(X)$,
 then the elliptic genus of the
conformal field theory only depends on the topology of $X$ and
hence we can speak  of $\chi(\tau,z;X)$
In this case    $\chi(\tau,z;X)$ can be interpreted as
 an    equivariant index of the Dirac operator
$\Dsl$ on the loop space $LX$. The parameter
$q$ accounts for rigid  rotations of a loop,
while $z$ accounts for rotations in the holomorphic
tangent space  $T^{1,0}X$  of the target.

We will be considering the elliptic genus for the case $X = {\rm Sym}^k(K3)$.
The elliptic genus for such $X$ is expressed in terms of the elliptic
genus of $K3$ itself. For any conformal field theory with Hilbert space $\CH$
we can consider the symmetric group orbifold of $\CH^{\otimes k}$.
Denote the Hilbert space of the orbifold theory by ${\rm Sym}^k(\CH)$. This has a
decomposition into twisted sectors given by
\eqn\lngstr{
\CH({\rm Sym}^k(\CH)) = \oplus_{\{ k_r \} } \otimes_{r>0} {\rm Sym}^{k_r}(\CH_r )
}
where the sum is over partitions of $k$:
\eqn\a{
\sum rk_r = k
}
The space $\CH_r$ is isomorphic to   the space $\CH$. It corresponds to
``strings of length $ 2\pi r$ '' where we scale the usual parameter
$\sigma \sim \sigma + 2\pi$ by a factor of $r$.    Thus configurations in the
symmetric product orbifold theory may be visualized as in \windingstrings.

\ifig\windingstrings{ A configuration of strings in the symmetric product
conformal field theory. }
{\epsfxsize2.0in\epsfbox{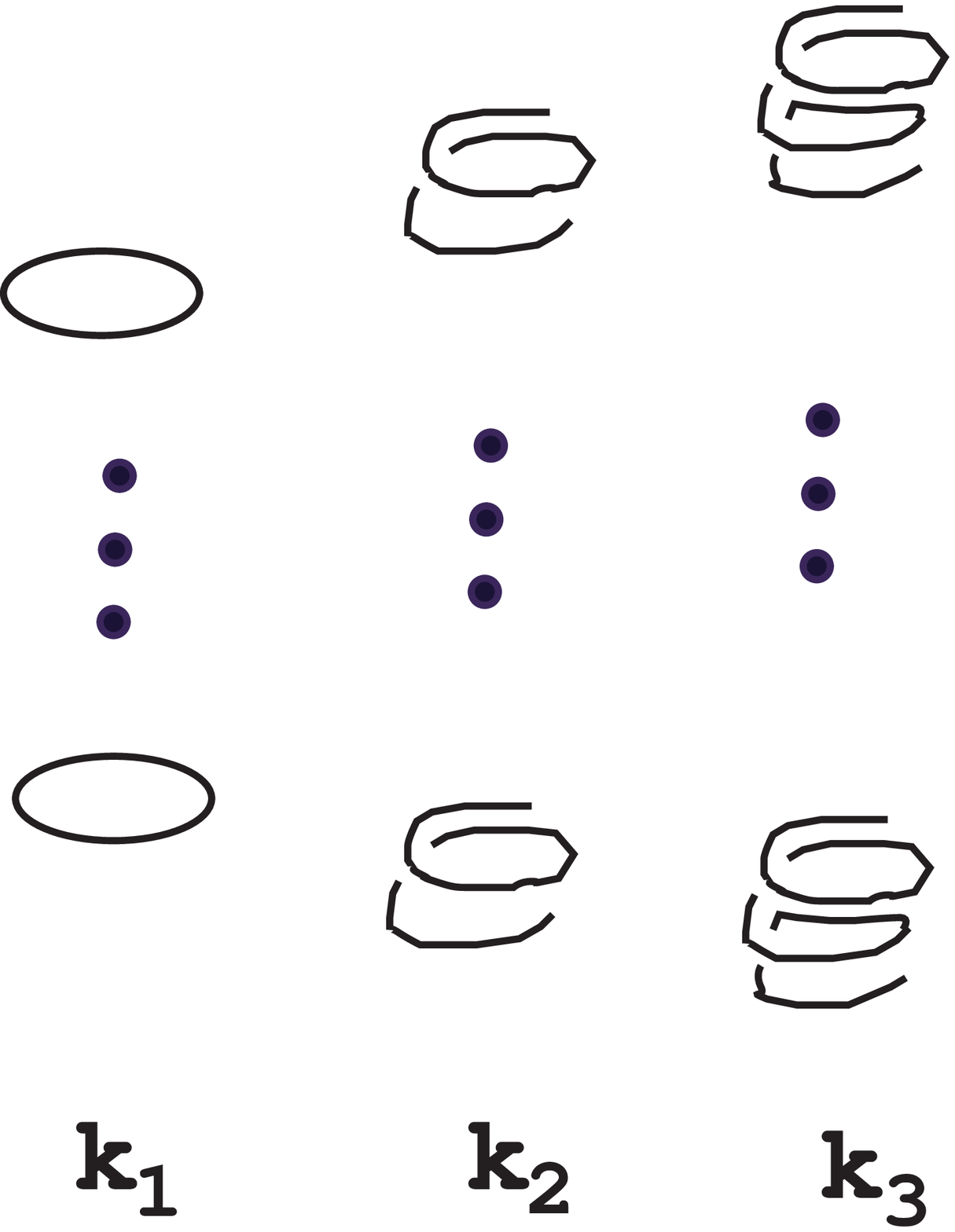}}

Now, if $\CH$ is a conformal field theory based on a sigma model with target
space $M$ then \lngstr\ implies an identity on the orbifold elliptic genus
for ${\rm Sym}^k(M)$. To be specific, if
\eqn\a{
\chi(\tau,z;M) = \sum c(n,\ell) q^n y^\ell
}
then \dmvv\
\eqn\ellgenfp{
 \sum_{k=0}^\infty p^k \chi({\rm Sym}^k M; q,y) =
\prod_{n>0, m\geq 0, r} {1 \over  (1-p^n q^m y^r)^{c(nm,r)}}
}

In the AdS/CFT correspondence we apply this to $M=K3$.
The elliptic genus of $K3$ can be computed (say, from orbifold limits or
Gepner models) and is
\eqn\kthree{
\chi(q,y;K3) = 8 \biggl( \bigl({\vartheta_2(z \vert \tau) \over
\vartheta_2(0\vert \tau) } \bigr)^2
+
\bigl({\vartheta_3(z \vert \tau) \over  \vartheta_3(0\vert \tau) } \bigr)^2
+
\bigl({\vartheta_4(z \vert \tau) \over  \vartheta_4(0\vert \tau) }
\bigr)^2\biggr)
}
and therefore, $\chi(\tau,z; {\rm Sym}^k(K3)) $ is explicitly known.

The decomposition in terms of theta functions is given by \KawaiTE
\eqn\thetachidec{ \chi(q,y;K3)  = h_0(\tau) \Theta_{0,1} + h_1(\tau)
\Theta_{1,1}}
 with
\eqn\achzero{ h_0(\tau) = \eta(\tau)^{-6} \biggl( 6 (\vartheta_2
\vartheta_4)^2\vartheta_3(2\tau) - 2 (\vartheta_4^4 - \vartheta_2^4)
\vartheta_2(2\tau)\biggr) }
\eqn\achone{ h_1(\tau) = \eta(\tau)^{-6} \biggl( 6 (\vartheta_2
\vartheta_4)^2\vartheta_2(2\tau) + 2 (\vartheta_4^4 - \vartheta_2^4)
\vartheta_3(2\tau)\biggr)
}%

with
\eqn\qexpach{ \eqalign{ h_0 &= 20 + 216 q + 1616 q^2 + \cdots\cr h_1
& =q^{-1/4} ( 2 - 128 q + \cdots )\cr} }

Many other interesting aspects of the elliptic genus of $K3$ and its
symmetric products, including relations to automorphic  infinite
products   can be found in \tkawai.

\subsec{Expressing the elliptic genus as a  Poincar\'e Series  }

Returning to our main theme, we will
explain the basic formula first in a simplified situation.
Then we state without proof the analogous result for weak Jacobi forms.
The proof may be found in \DijkgraafFQ.

Let $f \in M_w^*$ be a weak modular form for $SL(2, \IZ)$ of
weight $w\leq 0$. The adjective ``weak'' means that $f$ is allowed
to have a pole of finite order at the cusp at infinity, but no other singularities
in the upper half plane. Thus, the Fourier expansion of $f$ takes the form:
\eqn\a{
f(\tau) = \sum_{n\geq 0} D(n) q^{n+ \Delta}
}
We refer to the {\it finite} sum
\eqn\polarpar{
f^-(\tau) = \sum_{n+\Delta<0} D(n) q^{n+ \Delta}
}
as the {\it polar part}.

In the physical context,
$\Delta = -c/24$, for a unitary CFT, where  $c$ is the  central charge
of the Virasoro algebra. Moreover,
$w=-d/2$, where $d$ is the  number of noncompact bosons in the CFT.
Unfortunately, the letters $c,d$ are quite standard in the
theory of modular forms so there is a clash of conventional
notations. We will try to avoid the use of $c,d$ for central
charge and noncompact dimensions in what follows and use $\Delta, w$ instead.

It turns out to be essential to introduce a map
\eqn\bolop{M_w^* \to M_{2-w}^*}
 The explicit map is
\eqn\bolopii{
f(\tau) \rightarrow \CZ_f(\tau) := \bigl(q {\p \over  \p q}
\bigr)^{1-w} f
}
The fact that the right hand side of \bolopii\ is a modular form is
sometimes called Bol's identity. Note that in terms of the
Fourier expansion we have:
\eqn\bopiii{
\CZ_f = \sum_{n\geq 0 } \tilde D(n) q^{n+\Delta}
}
where
\eqn\bopiv{
\tilde D(n) = (n+ \Delta)^{1-w} D(n) .
}

Given a polynomial $\wp$ in $q^{-1}$ one can construct by hand
a modular form of weight $w$ by averaging over the modular group
to produce a Poincar\'e series
\eqn\poinser{
\sum_{\Gamma_\infty\backslash \Gamma} (c \tau + d)^{-w}
\wp({a \tau + b \over  c \tau + d} )
}
Note that we must sum over cosets of the stabilizer of $i\infty$, that is,
we sum over $\Gamma_\infty\backslash \Gamma$ where
\eqn\stabinf{
\Gamma_{\infty} := \{ \pmatrix{1 & \ell \cr 0 & 1 \cr}\vert \ell\in \IZ \}
}
The resulting sum is  convergent for $w>2$.

\lref\HarveyFQ{
J.~A.~Harvey and G.~W.~Moore,
``Algebras, BPS States, and Strings,''
Nucl.\ Phys.\ B {\bf 463}, 315 (1996)
[arXiv:hep-th/9510182].
}

In general, weak modular forms of positive weight $w>0$ are
not uniquely determined by their polar parts. If the space of
modular forms
$M_w$ is nonzero one can always add an nonzero element to
\poinser\ to produce another form with the same polar part.
However, if a form is in the image of
the map \bolopii\ then it is in fact completely determined by its
polar part. To see this, first note that
$\CZ_f$ has no constant term. Next we use a pairing between
weak modular forms and cusp forms which was quite useful in
\HarveyFQ. If $f\in M_w^*$ and $g\in S_w$ is a cusp form
then we can extend the Petersson inner product by
\eqn\pairing{
(f,g): = \lim_{\Lambda \to \infty}
\int_{\CF_\Lambda} {dx dy\over y^2} y^w f(x+iy )\overline{g(x+iy)}
}
Here $\CF_{\Lambda}$ is the intersection of the standard fundamental
domain of $PSL(2,\IZ)$ with the set of $\tau = x+iy$ with $y\leq
\Lambda$. Using integration by parts we can see that $\CZ_f$ is
orthogonal to   the space of cusp forms $S_{2-w}$, and hence it is
determined by its polar part.

Let us summarize:
We can reconstruct $\CZ_f$ from the polar part
\eqn\polpart{
\CZ_f^- = \CZ_{f^-}
= \sum_{n+\Delta<0} \tilde D(n) q^{n+\Delta}
}
(which is a {\it finite} sum)
  via
\eqn\recon{
 \CZ_f(\tau) =
\sum_{\Gamma_\infty\backslash \Gamma} (c \tau + d)^{w-2}
\CZ_f^-({a \tau + b \over  c \tau + d} )
}
This is the kind of formula we are going to interpret in
terms of AdS/CFT.

\subsubsec{Digression 2:  Rademacher's formula  }

In the next two subsections we pause to make two more small digressions
concerning some related issues:
Rademacher's formula,   Cardy's formula, and the
applications to black hole entropy.

The Rademacher formula
is a formula for the Fourier
coefficients of $f(\tau)$ which is particularly useful
for questions about the asymptotic nature of the Fourier
coefficients. The formula is easily derived from \recon\
   by taking  a Fourier
transform. On the left hand side we have:
\eqn\rdlhs{
\int_{\tau_0}^{\tau_0+1} e^{-2\pi i (\ell + \Delta) \tau } \CZ_{f}(\tau) d \tau = \tilde D(\ell)
}
on the right hand side, after a little manipulation  we have a sum of integrals of the form:
\eqn\rdrhs{
\int (c\tau+d)^{w-2} e^{-2\pi i (\ell + \Delta) \tau } e^{2\pi i (n+\Delta){a\tau + b \over c \tau+d}} d\tau
}
which can be expressed in terms of Bessel functions. The precise relation we find is
\eqn\rdfrmla{
\eqalign{
 & D(\ell) = 2 \pi \sum_{n+\Delta<0}
 \biggl({\ell +\Delta \over \vert n+ \Delta \vert}\biggr)^{(w-1)/2} D(n)
\cdot \cr
&
\cdot \sum_{c=1}^{\infty} {1 \over  c} Kl(\ell+\Delta,  n+\Delta;c)
   I_{1-w}\biggl({4 \pi \over  c}
\sqrt{\vert n+ \Delta\vert (\ell+ \Delta)} \biggr).  \cr}
}
where $I_\nu(x)$ is the Bessel function growing exponentially at $\infty$
\eqn\besselfn{
I_w(x) \sim {1\over \sqrt{2 \pi x} } e^x \qquad \Re(x) \to + \infty
}
while
\eqn\kloost{
Kl(n,m;c):=\sum_{d \in (\IZ/c\IZ)^*}
\exp\biggl[ 2 \pi i (d {n  \over  c} +  d^{-1}{m  \over  c})\biggr]
}
is a Kloosterman sum.

While \rdfrmla\ is a terribly complicated formula, it is in fact also very
useful since it gives the asymptotics of Fourier coefficients of
modular forms for large $\ell$. In fact, it can be a very efficient
way to compute  the Fourier coefficients exactly if they are known,
for example, to be integral.

In the physics literature the
 leading term,
\eqn\crdyfrm{
D(\ell) \sim  {D(0)\over \sqrt{2}}
\biggl( {  ( \ell +\Delta )^{\half w - {3\over 4} }
\over \vert   \Delta \vert^{\half w-{1\over 4} }   }\biggr)
 \exp\left[ 4\pi \sqrt{ \vert \Delta\vert (\ell + \Delta) } \right]
}
 is known as ``Cardy's formula.''  It gives the
  ``entropy of states at level $\ell$''

The subleading exponential corrections are organized in a beautiful
way by Farey sequences. See  \refs{\rademacheri,\rademacherii,\apostol}
or \DijkgraafFQ, appendix B for details.

%
%

\subsubsec{Digression 3: Black hole entropy}

One very striking application of Cardy's formula in the
string literature is to the statistical accounting for
the entropy of certain special black holes. This was first
proposed in a famous paper of A. Strominger and C. Vafa \sv.

As we have mentioned, the spacetime
$AdS_3 \times S^3 \times K3$ is obtained as a near-horizon
geometry from a limit of
a system of $Q_1$ $D1$-branes and $Q_5$ $D5$-branes wrapping
$S^1 \times K3$.
The   ``BPS states'' of this system of branes correspond to special
black hole solutions of 5-dimensional supergravity.
The black hole solution is characterized by three charges
$Q_1, Q_5, N$. In the D-brane system, $Q_1, Q_5, N$ specify quantum numbers of
BPS states;   there is a $\IZ_2$-graded  {\it vector space}
of such states: $\CH^{BPS}_{\gamma}$,
with charges $\gamma = (Q_1, Q_5, N)$.
The elliptic genus counts the super-dimension of these vector spaces of BPS states:
\eqn\sydmis{
\chi(q, {\rm Sym}^k K3) = \sum q^N
{\rm sdim} \CH^{BPS}_{\gamma=(Q_1,Q_5, N)}
}
The Cardy formula  then gives:
\eqn\crdy{
I \sim \exp\left( 2\pi \sqrt{Q_1 Q_5 N} \right)
}
and confirms the supergravity computation of the Beckenstein-Hawking entropy \sv.
\foot{It is important to bear in mind that this is actually counting with
signs. It is counting vectormultiplets minus hypermultiplets, and can lead to
cancellations, and hence it can underestimate   the entropy. In the case examined in
\sv\ it gives the ``right'' answer, i.e. the answer that coincides with
supergravity. }

The Rademacher formula gives an infinite series of subleading corrections
\eqn\sublead{
 \sim \exp\left(  {2\pi \over c} \sqrt{Q_1 Q_5 N} \right) \qquad c=2,3,4,\dots
}
 organized by  terms in the Farey sequences. In     section 2.6 we will
discuss  the physical interpretation of these subleading corrections.

\subsubsec{ Poincar\'e Series for the Elliptic Genus}

Finally, let us return to the main task of this section: Expressing the
elliptic genus as a Poincar\'e series in a form suitable to interpretation
within the AdS/CFT correspondence.

The manipulations of section 2.5 above have analogs for Jacobi
forms. Let $J_{w,k}$ denote the space of weak Jacobi forms of weight
$w$ and index $k$.
%
%
 The analog of the polar part
\polarpar\ is the sum over Fourier coefficients with
\eqn\jpolarpart{ 4kn - \ell^2 <0 . }

Applied to the elliptic genus the relevant Poincar\'e series becomes:
\eqn\egps{
\eqalign{
 & \CZ_\chi(\tau,z) =
2 \pi  \sum_{(\Gamma_{\infty} \backslash \Gamma)_0 }
  \sum'_{m,\mu }
\tilde c_\mu\bigl( 4km-\mu^2; {\rm Sym}^k(K3)\bigr)
 \qquad \qquad \cr
&
  \exp[- 2\pi i k {c z^2 \over  c \tau +d}]
\Theta^+_{\mu,k}({z \over  c \tau +d} ,
{a \tau + b \over  c \tau + d}  )\cr\
&
\bigl(  c \tau + d  \bigr)^{-3}
\exp\biggl[2 \pi i \bigl( m -{ \mu^2\over  4k}\bigr)
 {a \tau + b
\over  c \tau + d} \biggr]\cr } } where $(\Gamma_{\infty} \backslash
\Gamma)_0$ is the sum over relatively prime pairs $(c,d)$ with
$c\geq 0$, while $ \sum'_{m,\mu}$ is a finite sum over $(m,\mu)$
with $4km-\mu^2 < 0$, and $\Theta^+_{\mu,k}$ was defined in \thetmk.

In the next section we
are   going to sketch how this sum can be interpreted as a
sum over solutions to 10D supergravity.

\bigskip
{\it Note added, Dec. 8, 2007}: Don Zagier pointed out an important
error in versions 1-3 of this paper. The map
\eqn\ft{ \phi = \sum c(n,\ell) q^n y^\ell \to \tilde \phi  = \sum
\tilde c(n,\ell) q^n y^\ell } with \eqn\jscd{ \tilde c(n,\ell) =
\vert n- \ell^2/4k\vert^{3/2-w} c(n,\ell) }
does {\it not} map Jacobi forms $J_{w,k} \to J_{3-w,k}$, contrary to
what was asserted in versions 1-3. Nevertheless, for
$n-\ell^2/4k>0$, the $\tilde c(n,\ell)$ can be obtained as Fourier
coefficients from the Poincar\'e series \egps. For further details
see the corrected version 3 of \DijkgraafFQ, as well as \ManschotHA,
which writes a regularized Poincar\'e series for the elliptic genus
itself, and not its ``Fareytail transform.''

\subsec{  AdS/CFT Interpretation of the Poincar\'e Series}

In the previous section we wrote down the Poincar\'e series \egps\ for
the elliptic genus. This is a mathematical fact, and we
are regarding this exact result as a precious piece of
``experimental data'' to tell us how to formulate the
string theory side of the AdS/CFT correspondence.
As we will see, the precise formulation of string theory on
$AdS_3 \times S^3 \times K3$ is full of interesting subtleties.
We will now proceed to interpret the various factors in \egps\ in
physical terms.

\subsubsec{Average over $\Gamma_\infty\backslash \Gamma$ and  BTZ black holes}

We are going to describe the AdS dual to a conformal field theory computation
of a partition function. Therefore, the conformal boundary of the $AdS_3$ should
be a torus. Therefore, we will be looking at 3-dimensional geometries filling in
$S^1_\phi \times S^1_t$.
The metric will accordingly have boundary conditions:
\eqn\mtrcbdry{ds^2 \rightarrow r^2 \vert d \phi + i d t \vert^2 + {dr^2 \over
 r^2}
 }
for $r\to \infty$. Here
$(\phi + i t) \sim
(\phi + i t)  + 2 \pi (n + m \tau)$, $n,m\in \IZ$, and $\tau $ determines the
conformal structure of the torus at infinity.

The only smooth complete hyperbolic geometry satisfying these conditions has the
topology of a solid torus. One way to realize this geometry is to take a
quotient of the upper half plane
$\IH = {\bf C } \times \IR^+$ by the group $\IZ$ acting as
$(z,y) \to (q^n z, \vert q^n \vert y)$. We can compactify the
space by adding the boundary at infinity $\IC^*$. We must omit $0,\infty\in \hat \IC$
to get a properly discontinuous group action.
Topologically, the resulting space is a solid torus.
%
%

While the hyperbolic geometry is unique, in order to do physics we need
to make a choice of what is called ``space'' and what is called  ``time''
in the torus at infinity. This choice will affect computations of action, entropy
etc. It is this choice which accounts for the sum over $\Gamma_\infty\backslash\Gamma$,
that is, over relatively prime
integers $(c,d)$ in \egps. Geometrically, $(c,d)$ describes the unique primitive
homology cycle which becomes contractible upon filling in the torus with a
solid torus.

For example, let us choose coordinates $(\phi,t)$ on
$S^1 \times S^1$. If   we choose the term
$(c=0,d=1)$ then it is the  ``spatial'' $\phi$-circle which is filled in.
In this case the geometry has the interpretation of an  ``AdS gas'' -- that is,
we analytically continue the time in Lorentzian AdS and identify it
with $t_E \sim t_E + \beta$.

On the other hand, in the term corresponding to
$(c=1,d=0)$  it is Euclidean  ``time'' - the  $t$-circle - which  is filled in.
In this case we have the Euclidean  ``BTZ black hole.'' Note that the spatial
circle is noncontractible: There is a hole in space, and it is in fact
correctly interpreted as a true black hole solution of gravity,
as shown in great detail in
\btz\carlipbook.

The general solution is labelled by a point in
\eqn\queehat{\Gamma_{\infty} \backslash \Gamma \cong \hat {\IQ}
}
and is labelled by the homology class of the primitive cycle
which is contractible.
This family of black holes is the proper interpretation of
what Maldacena and Strominger
termed an ``$SL(2,Z)$ family of black holes'' in
\stringexclusion.  Thus, the first, and most basic aspect of
\egps\  is that it is a sum over this family of black holes
(including  the AdS gas $(c=0,d=1)$).
\foot{An heuristic version of this sum was first written down in
\stringexclusion.}

\subsubsec{Low energy Chern-Simons theory }

Now, we would like to compute the contribution of the string theory path integral to
each term in the sum over pairs $(c,d)$ in \egps.
A crucial point is that the elliptic genus is unchanged under
deformation of parameters. This allows us to focus on the low
energy and long-distance limit of the reduction of 10d supergravity on
 $AdS_3 \times S^3 \times K3$. In this limit, the dominant term in
the supergravity action is that of a Chern-Simons theory. The
Chern-Simons supergroup is  \deboer\
\eqn\a{
SU(2\vert 1,1) \times SU(2\vert 1,1)
}
and the explicit action is
\eqn\a{
{k \over 4\pi} \int {\Tr} (\CA d \CA + {2\over 3} \CA^3 ) - {\Tr} (\CB d \CB + {2\over 3} \CB^3 )
}
The
$SU(1,1)\times SU(1,1)$ connections  are derived from the
negative curvature metric via $A_\pm \sim w \pm e$ where $w$ is the spin connection and
$e$ is the dreibrein \WittenHC \carlipbook. The  $SU(2)\times SU(2)   $ gauge fields
arise from Kaluza-Klein reduction on  $S^3$. For a detailed derivation
of these terms in the action see \degersezgin\LuUW\ArutyunovBY.

We must choose boundary conditions for the Chern-Simons gauge fields.
The boundary values of the connections for
$
SU(2\vert 1,1)_L $,  and $ SU(2\vert 1,1)_R
$
couple to CFT left- and right-movers, respectively. The boundary conditions
\mtrcbdry\ determine boundary conditions on the $SU(1,1)$ gauge fields.
In addition: The $SU(2)$ gauge fields become {\it flat} at infinity
and the proper boundary conditions are:
\eqn\a{
 A_u du  \rightarrow {\pi \over 2 \Im \tau} z \sigma^3 du
}
where $u  = i ( \phi + it ) /(c \tau + d) $

Because of our choice of fermion spin structures the boundary conditions
of the right-moving $SU(2)$ gauge fields should drop out. This point deserves to be
understood more fully.

\subsubsec{  Spinning in 6-dimensions }

Actually, we have not yet fully enumerated the distinct types of
geometry that we must sum over. When we include the $z$-dependence in
the elliptic genus it is necessary to consider {\it six-dimensional geometries}.
This leads to an interpretation of the sum  on $\mu$ in \egps.

The BTZ black holes have natural generalizations
to quotients of the form
\eqn\a{
\IZ \backslash (\IH^3 \times S^3)
}
with $\IZ$ acting on $S^3 = SU(2)$ by
\eqn\a{
U \to \tilde U = e^{- i {\mu\over 2k} (t + \phi) \sigma^3} U
}
These correspond to solutions spinning in six dimensions with
$2j_L  = \mu$. Such solutions
have been nicely described in detail in \cvl. Closely related
smooth solutions associated with BPS states have been
described in \LuninIZ.

In the effective $SU(2)$ Chern-Simons theory these solutions
correspond to the insertion of a Wilson line in the center of
the solid torus as in \wilsonline. Since the $SU(2)$ theory is governed by a Chern-Simons
theory we expect to see the wavefunction associated to such theories in the
partition function.
It is well-known that these wavefunctions are given by the affine
Lie algebra characters of $SU(2)$ level $k$ current algebra for
spin $j$. Another basis of wavefunctions count states at definite
values of $J_0^3$. These are given by level $k$ theta functions:
\eqn\a{
 \exp[- 2\pi i k {c z^2 \over  c \tau +d}]
\Theta^+_{\mu,k}({z\over  c \tau +d} ,
{a \tau + b \over  c \tau + d}  ).
}
%

\ifig\wilsonline{ A black hole spinning in 6 dimensions is effectively
equivalent to the partition function on a solid torus with a Wilson line insertion. }
{\epsfxsize2.0in\epsfbox{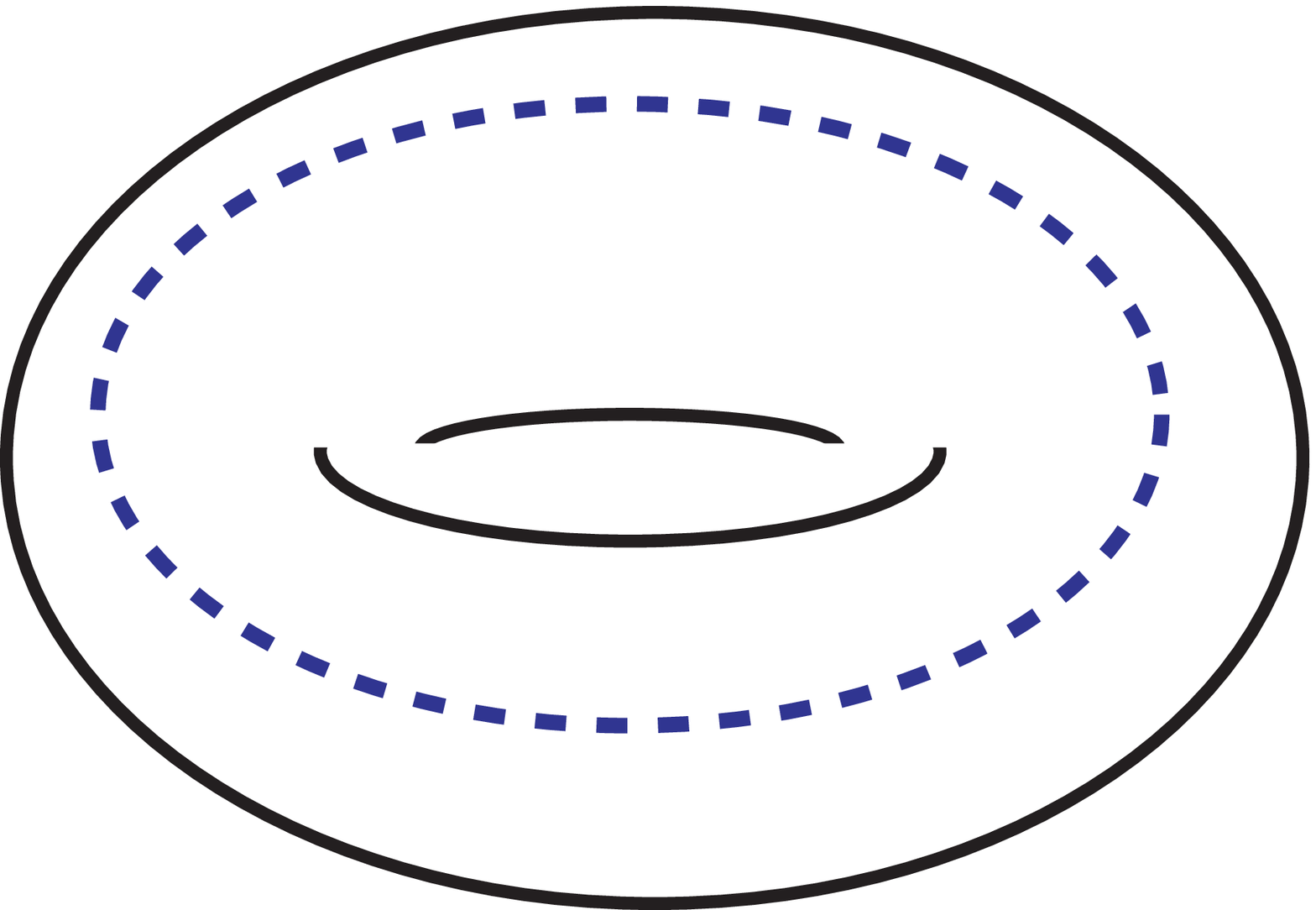}}

To summarize,   we can interpret the contribution of $(c,d)$ and $\mu$ as
a BTZ black hole with homology class $(c,d)$ contractible and with Wilson
lines inserted so that the Chern-Simons wavefunction has definite values of
$\mu$ modulo $k$, as in \wilsonline.

\subsubsec{The light particles of supergravity  }

Let us now interpret the sum over the polar part in \egps,
\eqn\a{ \sum_{m: 4km-\mu^2 < 0}
 \tilde c_\mu\bigl( 4km-\mu^2; {\rm Sym}^k(K3)\bigr)
}
In order to do this we must address some aspects
of the Lorentzian version of the AdS/CFT correspondence.

In the Lorentzian version, there is an isomorphism of Hilbert spaces
between the Hilbert space of the boundary conformal field theory and some
much more mysterious Hilbert space of quantum gravity (string theory)
on some interior space.  The Hilbert space of the conformal field theory is
rather well-understood. We will view it as a  Hilbert space  graded by
the values of $(L_0,J_0)$. In the elliptic genus, the left-moving Ramond sector
states have quantum numbers $(m,\mu)$ which we identify as  the
eigenvalues
$$(m,\mu)= (L_0- c/24,J_0)$$

Now, we expect such states to correspond to states in the quantum gravity
Hilbert space. Symmetry principles (i.e. matching of superconformal symmetries)
show us that
we must interpret $L_0$ as the 2+1 dimensional energy + spin, while $J_0$ should
be viewed as the $J^3$ eigenvalue for spin in the $S^3$ directions.

\lref\StromingerEQ{
A.~Strominger,
``Black hole entropy from near-horizon microstates,''
JHEP {\bf 9802}, 009 (1998)
[arXiv:hep-th/9712251].
}

{}From the point of view of quantum gravity, there is an important distinction
between states which are small perturbations on an AdS background - we will
refer to these as ``particle states'' -  and states which form black holes.
The distinction is governed by the ``cosmic censorship bound''
\breckenridge\cveticyoum\cvl. Black holes correspond to semiclassical
states in $\CH_{\rm string}$. The corresponding states in
$\CH_{CFT}$ have $L_0$ in the Ramond sector related to the mass $M$
of the black hole by $M=L_0-c/24$ \StromingerEQ. On the other hand,
the condition for a black hole to have a nonsingular horizon is
$4kM-J_0^2\geq 0$ \breckenridge\cveticyoum\cvl.  Such states therefore
have $4km-J_0^2 \geq 0$. Thus the unitarity region in the
  $(m,\mu)=(L_0-k/4, J_0)$ plane is divided into two regions: Supergravity
states with $-k^2\leq 4km-\mu^2< 0$ are not sufficiently massive to
form black holes, corresponding to the shaded region in \rstates,
while states with $4km-\mu^2\geq 0$ will form black holes.  Thus,
{\it the states which do not form black holes correspond precisely
to the  to the polar part of the Jacobi form! } Moreover, the
degeneracy $ c_\mu\bigl( 4km-\mu^2; {\rm Sym}^k(K3)\bigr)$ is
precisely that of right-BPS supergravity particles from Kaluza-Klein
reduction of $(2,0)$ supergravity on $AdS_3 \times S^3$ \deboer.

\ifig\rstates{ The states in the shaded region are not sufficiently energetic
to form black holes. These states have quantum numbers corresponding to the
polar part of the elliptic genus. Note that quantum numbers not on the $\ell = 2J_0^3$
axis are {\it not} BPS states. The discussion above pertains to states which are
{\it right}-BPS. }
{\epsfxsize2.5in\epsfbox{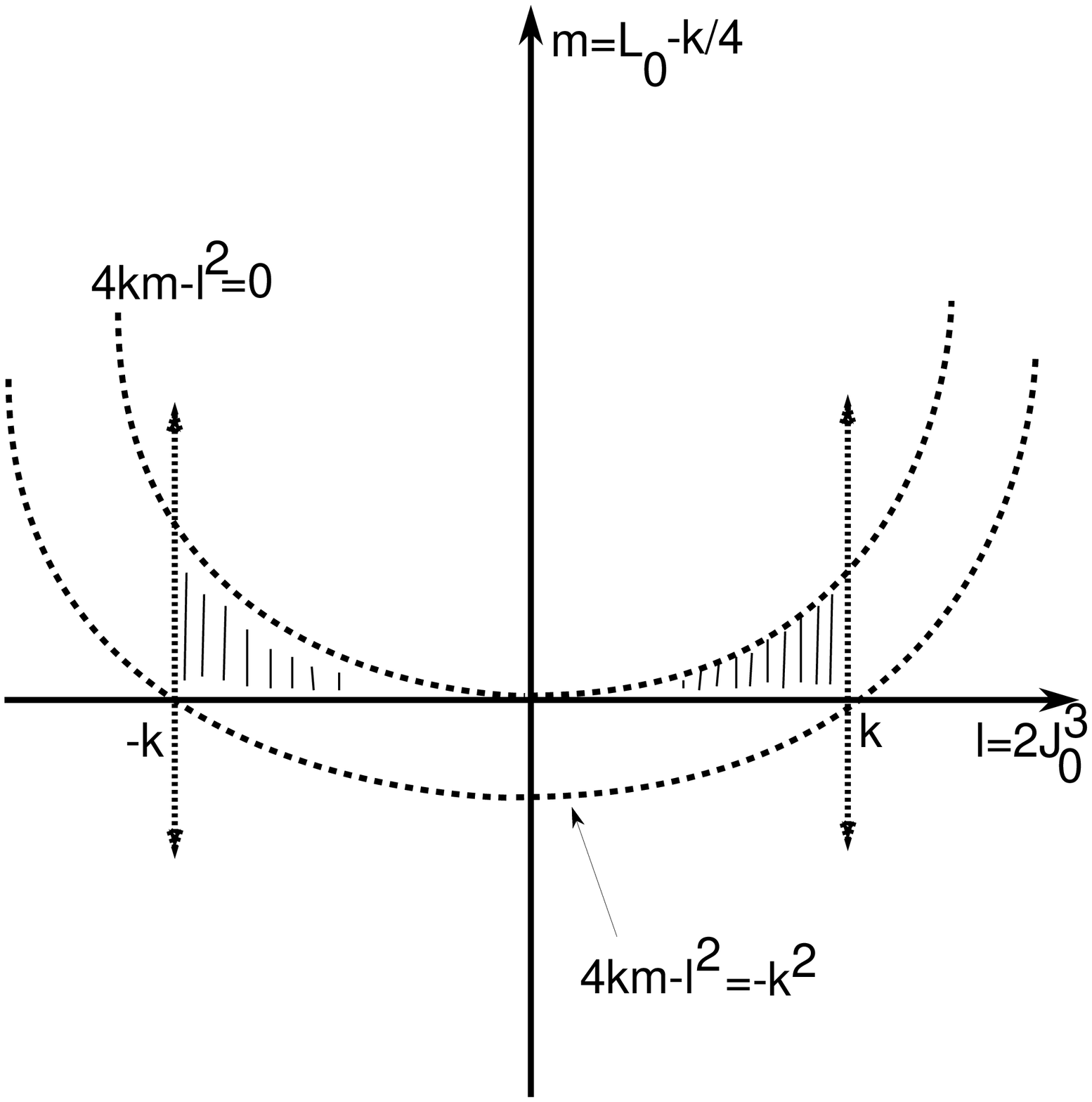}}

\subsubsec{ Gravitational action and  final factor}

According to our interpretation, the final  factors
\eqn\finfact{
\bigl(  c \tau + d  \bigr)^{-3}
\exp\biggl[2 \pi i \bigl( m -{ \mu^2\over  4k}\bigr)
 {a \tau + b
\over  c
\tau + d} \biggr]
}
should arise from a careful evaluation of an analytic continuation
of $SU(1,1) \times SU(1,1)$ Chern-Simons theory to Euclidean
signature.

Thus one is naturally let to attempt a careful
evaluation of the gravitational action for the spinning
extremal black holes. The Einstein action is
\eqn\a{
{1\over 16 \pi G} \int  \sqrt{g} (\CR - \Lambda) +
{1\over 8 \pi G} \int  K
}
where $K$ is the second fundamental form of the boundary.
Since the Einstein action on $AdS$ is infinite it must be regularized.
The standard way to do this is to introduce a boundary, thus necessitating
the second term. The difference of such actions between two geometries
in the family \queehat\  can be evaluated in a well-defined
way and gives:
\eqn\diffr{
\pi k \left( \Im \tau - \Im ({a \tau + b \over c \tau + d} ) \right)
}
Moreover, the computations of \cvl\ produce such an entropy factor
weighted by $m-\mu^2/4k$ in the six-dimensional case.

Upon taking a $\bar \tau \to \infty $ limit the expression \diffr\  closely resembles
\finfact, but, so far as we know, there is no honest and convincing derivation of
\finfact\ in the literature starting from the Chern-Simons approach.

Note that \finfact\ is {\it odd} under $(c,d) \to (-c,-d)$. This is
the reason we must put a restriction $c \geq 0$ on the Poincar\'e
series \egps. The transformation $(c,d) \to (-c,-d)$ corresponds to
the diffeomorphism $-1$ on the boundary torus. The fact that the
summand in \egps\ should be understood better. Perhaps it is due to
the fact that only Ramond groundstates contribute.

 \subsec{ Summary: Lessons \&  Enigmas}

We have presented some evidence to suggest that the   full AdS-interpretation of the
elliptic genus of the boundary conformal field theory can be expressed in the form
\eqn\lessone{
Z_{\chi} = \sum \Psi^{CS}_{SU(2\vert 1,1) }
}
where $\Psi^{CS}_{SU(2\vert 1,1) }$ is a wavefunction for a Chern-Simons theory and
where the sum is over Euclidean solutions of supergravity
of spinning black holes with supergravity particles in $AdS_3 \times S^3$.
It should be clear to the reader that there are gaps and enigmas in this
story. For examples,

\item{1.} Why do we need to take the Serre dual to get a reasonable formula?

\item{2.}
What is the origin of the factor
 \eqn\convfact{
1/(c\tau + d)^3
}
from the string partition function? Note that this factor is crucial for the
convergence of the sum over $(c,d)$. It also has the pleasant property that
$Z_\chi dz \wedge d \tau$ is a well-defined half-density on the universal
elliptic curve.

\item{3.} Is it sufficient to focus purely on the Chern-Simons sector to
evaluate the path integral or must one take into account the full tower of
string fields? (We have been assuming the latter contribute a
trivial factor to $Z_\chi$, because of its topological nature.)

\item{4.}  Perhaps the most important enigma is the origin of the sum over
the polar part in \egps. This is probably saying something significant
about  the Hilbert space of quantum gravity.
It indicates that the nature of the isomorphism between the CFT Hilbert
space and the string theory Hilbert space is qualitatively different
for the   infinite set of conformal field theory states above the cosmic
censorship bound.
What replaces a sum over states in the Euclidean quantum gravity Hilbert space
is a sum over a special set of geometries. Note in particular that the
$(m=0,\ell=0)$ term does {\it not } contribute. These are the unique quantum
numbers (the so-called ``$M=0$ BTZ'' black hole) of states which are simultaneously
topological and black holes. It is possible that this structure is related to
 the phenomenon of ``asymptotic darkness'' that has been advocated by
T. Banks \BanksIA\BanksVP.

\subsec{Applications}

Whether or not one believes the physical interpretation advocated in the previous
section, the formula \egps\ is true, and has some    some nice applications.

One  application is to the thermodynamics of string theory on Euclidean
$AdS_3 \times S^3 \times K3$.  One discovers
a   3-dimensional version of the deconfining phase transition of large $N$
$\CN=4$ Yang-Mills theory discussed by Witten \wittenads. In the $AdS_3$
case one   studies the   partition function as a function of
\eqn\a{
\tau = \Omega + i \beta
}
where $\Omega$ is the spin fugacity and $\beta$ is the inverse temperature.
In the large $k$ limit $Z_\chi$ becomes a piecewise analytic function
of $\tau$.  It is simplest to study the partition function
in the $(NS,R)$ sector (by setting $z=-\tau/2$).   As $k \to \infty$ at fixed
$\tau$ the
dominant geometry is characterized by the pair
$(c,d)$ which maximizes
\eqn\a{
{ {\rm Im } \tau \over \vert c \tau + d \vert^2}
}
This geometry contributes a term of order
\eqn\ordercont{
{1\over \vert c \tau + d \vert^3} \vert \tilde c(-k^2) \vert \exp\biggl[ {\pi k \over 2}
{ {\rm Im } \tau \over \vert c \tau + d \vert^2} \biggr]
}
The standard keyhole region fundamental domain $\CF$ for $SL(2,\IZ)$ has the property
that the modular image of any point $\tau\in \CF$ has an imaginary part
${\rm Im} \tau' \leq {\rm Im}\tau$. Therefore, the phase domains are given by
$\cup_{n\in \IZ} T^n \cdot \CF = \Gamma_\infty\cdot \CF$ and its modular images.

\lref\MaldacenaKR{
J.~M.~Maldacena,
``Eternal black holes in Anti-de-Sitter,''
arXiv:hep-th/0106112.
}

As a second application we note that a computation similar in spirit to what
we have discussed was performed by Maldacena to resolve a sharp version of the
``black hole information paradox'' for eternal AdS black holes.
See \MaldacenaKR.

\subsec{Speculations on future applications of AdS/CFT to number theory}

In this section we present some speculations on ways in
which the AdS/CFT correspondence might have some interesting
interactions with number theory. Our speculations are based on
ongoing discussions with A. Strominger, and have at times involved
B. Mazur, and S. Gukov.
For some related ideas see \ManinHN. (Some overlapping remarks
were made recently in \BalasubramanianKQ\MaldacenaRF.)

\subsubsec{ Quotients of AdS/CFT   }

Suppose string theory  on $AdS_{n+1} \times K$ is dual to a conformal
field theory $\CC$. Suppose
\eqn\a{\Gamma\subset SO(1,n+1)\qquad or \qquad \Gamma\subset SO(2,n)}
is an infinite discrete group. Since $\Gamma$ acts as a group
of isometries in the bulk theory,
we can consider string theory on
\eqn\quotspace{
\Gamma \backslash (AdS \times K)
}
It is natural to ask if string theory on \quotspace\ makes sense, and
if so, whether it is
dual to some kind of ``quotient'' of  the conformal field theory  $\CC$
by $\Gamma$. Note that such a quotient, if it even exists,
 is very different from an orbifold of a
conformal field theory, for $\Gamma$ acts by conformal transformations on
the ``worldsheet'' rather than the ``target space'' of $\CC$.

Such a duality, if it were to make sense,  would have very interesting implications
in at least two ways.
First, there would be important applications to questions of
cosmology and time dependence in string theory. Second -- and more
central to the theme of these lectures -- there would be interesting
applications to number theory.
In the following sections we will sketch some of the possible applications.

The reader should be warned at the outset that there
 are nontrivial difficulties with the idea that AdS/CFT duality
can survive general quotients by such groups $\Gamma$.
The difficulties stem from the fact that the ``interesting'' groups
we wish to consider act on the conformal boundary at infinity,
$\p \IH^n$, but the action is sometimes ergodic. More precisely,
the boundary is divided into a disjoint union of two regions:
\eqn\domaindisc{
\p \IH^n = \Omega_{\Gamma} \cup \Lambda_{\Gamma}
}
The first region $ \Omega_{\Gamma}$ is the domain of discontinuity.
Here the group acts propertly discontinuously and the quotient
$\Omega_{\Gamma}/\Gamma$ is, for $n=2$, a Riemann surface. Note that this
Riemann surface can have cusps and several connected components.
The complementary region $\Lambda_{\Gamma}$ is called the limit set.
It is the closure of the set of accumulation points of $\Gamma$,
and the action on $\Lambda_{\Gamma}$ is ergodic. This means that any ``quotient''
of the boundary conformal field theory is going to have strange
behavior on $\Lambda_{\Gamma}$. To take an extreme example, there
are groups $\Gamma$ with no domain of discontinuity. Then the
classical quotient $\IH^n/\Gamma$ is a {\it compact} hyperbolic
manifold. So the ``boundary theory,'' if it exists, must surely
be something truly unusual.

In fact, the quotient by $\Gamma$   can produce strange causal structure
in the Lorentzian case,
a fact which probably indicates   large backreaction in the context of
supergravity. A related point is that the distance between image points
 $d(x,\gamma\cdot x)$ can get small, again indicating breakdown of
 the sugra approximation.  Indeed, the existence of a boundary
theory for groups $\Gamma$ with nontrivial limit set has been
argued against by Martinec and McElgin   \MartinecCF\MartinecXQ.

Nevertheless, a successful outcome would undoubtedly lead to many
very fascinating things, so let us suppose that a dual boundary
theory does exist
and briefly ask what it might be good for.

\subsubsec{  String Cosmology}

A few years ago, in \HorowitzXK, interesting cosmologies with singularities
were considered based on spacetimes of the form \quotspace.

\lref\CornalbaKD{
L.~Cornalba and M.~S.~Costa,
``Time-dependent orbifolds and string cosmology,''
arXiv:hep-th/0310099.
}

More recently, string theory with time-dependent singularities in
``soluble'' string models has come under some scrutiny. Amongst the
many investigations in this area is the work in \LiuFT\LiuKB\LiuYD\CornalbaKD\
which studies the $\IZ$-orbifold
of $\IR^{1,2}$ defined by the action
\eqn\parborb{
X:=
\pmatrix{ x^+ \cr x^{~}\cr x^-\cr} \quad \rightarrow\qquad g_0^n\cdot X = \pmatrix{ x^+ \cr x + n v x^+  \cr
x^- + n v x + \half n^2 v^2 x^+ \cr}
}
where $(x^+,x,x^-)$ are light-cone coordinates.
It turns out that  string perturbation theory in such backgrounds is highly problematic.
The  difficulties are expected to be a generic feature of strings in
cosmological singularities. Moreover, nonperturbative effects involving
 black holes are expected to be important \HorowitzMW.
This is relevant to the present discussion for the following reason.
Recall that $AdS_{1,2}$ is the universal covering space $  \widetilde{SL(2,R)}$.
The Lie algebra $sl(2,\IR) =\IR^{1,2}$ is Minkowski space.
 Consider the action on $AdS_{1,2}$ by $\IZ$ with
\eqn\a{
g \to g_0 g g_0^{-1},
}
where $g_0$ is a parabolic element. In
 the scaling region of $g=1$ these look like the cosmological models \parborb.
On the other hand, since there is a boundary theory summarizing all the nonperturbative
physics, it is reasonable to think, {\it provided the AdS/CFT correspondence
survives the quotient construction}, that the boundary theory contains some clue
as to the resolution of the cosmological singularity. Some investigations along these
lines were carried out in \KrausIV, but there is much more to understand.

%
%
%
%

\subsubsec{ Potential Applications to Number Theory: Euclidean version }

One of the possible applications of these ideas to number theory
concerns the theory of modular symbols.

Let us recall (in caricature)
 the $AdS/CFT$ computation of the 2point function
of spinless primary fields.
In AdS the tree-level 2-point function of scalar fields $\phi$ is
the Green's function:
\eqn\gfi{
(\Delta_1 + m^2) G(P_1,P_2) = \delta(P_1,P_2)
}
In $\IH^3$ we have the simple explicit formula:
\eqn\gfii{
G(P_1, P_2) = {1\over 2\pi} {e^{-2h d(1,2)} \over 1-e^{-d(1,2)} }
}
where
\eqn\gfiii{
\cosh d(1,2) = 1+ {\vert z_1 - z_2 \vert^2 + (y_1 - y_2)^2 \over 2 y_1 y_2} \qquad
 m^2 = 2h (2h-2)
}
One extracts the 2point correlator from the boundary behavior of the
Green's function:
\eqn\gfiv{
G(1,2) \to y_1^{2h} y_2^{2h} \langle \Phi_\phi(z_1) \Phi_\phi(z_2) \rangle \qquad
}
as $y_1,y_2\to 0$. This leads to the familiar result:
\eqn\gfv{
 \langle \Phi_\phi(z_1) \Phi_\phi(z_2) \rangle = {1\over \vert (z_1 - z_2)^{2h} \vert^2}
}
where $\Phi_\phi$ is the dual operator of \dualop.

Now, let $\Gamma \subset PSL(2,{\bf C} )$ be discrete and suppose AdS/CFT
``commutes with orbifolding.''
In the tree-level approximation, the
 Green's function on $\Gamma\backslash \IH^3$ is obtained by the
method of images. Therefore, according to \gfiv\ the boundary CFT
correlator should be obtained from the method of images. For a
primary field (with spin ) of weights $(h,0)$ this would lead to
\eqn\mthdimrgs{
\langle \Phi(z_1) \Phi(z_2) \rangle_{\Gamma\backslash \Omega_\Gamma }
 = \sum_{\Gamma} {1\over (z_1 - \gamma\cdot z_2)^{2h} } {1\over (cz_1 + d)^{2h}} .
}
We would like to stress that in general in CFT it is {\it not} true that
the conformal correlators on Riemann surfaces $\Gamma\backslash\Omega_{\Gamma}$ are
obtained by the method of images. While it is true that the Green's function of
a scalar field is obtained by summing over images, in the presence of
interactions there are further correlations between a source and its image point.
\foot{As a simple example, if $\phi$ is a free massless scalar field then $\langle \phi(1)\phi(2)\rangle$
is a sum of images, and therefore $\langle e^{i p \phi}(1) e^{-i p \phi}(2)\rangle$
is a {\it product} over images!}
Therefore, at best \mthdimrgs\  can apply in the large $k$
approximation (which justifies the tree-level supergravity). Even there,
AdS/CFT is making a highly nontrivial prediction for the boundary CFT
correlators.

Nevertheless, let us accept \mthdimrgs. Now  suppose
there is a flat gauge field in the low energy supergravity
 coupling to charged scalars $\phi^\pm$. Then
the boundary correlator becomes:
\eqn\mdsymgf{
\langle \Phi^+(z_1) \Phi^-(z_2) \rangle_{\Gamma\backslash \Omega_\Gamma }
 = \sum_{\Gamma} {e^{i q \oint_{\gamma} A}
\over (z_1 - \gamma\cdot z_2)^{2h} } {1\over (cz_1 + d)^{2h}}
}
For example, we could take $\Gamma = \Gamma_0(N)$ and  $A = f(z) dz$, for $f\in S_2(\Gamma_0(N))$,
a cusp form of weight 2.
In this way we obtain generating functions for modular symbols.
Curiously, functions very closely related to \mdsymgf\ have recently been studied in
attempts to understand the distribution of modular symbols \petridis. In view of this, it is
interesting to ask if  AdS/CFT could give new insights into questions involving modular symbols.

It is also natural to ask about nonabelian generalizations of \mdsymgf.
These can be written down. Recalling the relation between boundary
CFT and the Chern-Simons-Witten theory, one is lead to a new interpretation
of the Verlinde operators of that theory in terms of what might be
called ``quantum nonabelian modular symbols.'' We hope to describe this
in detail elsewhere.

\subsubsec{ Potential Applications to Number Theory: Lorentzian version}

As a second illustration of how applications to number theory might
arise, let us suppose  the Lorentzian AdS/CFT correspondence commutes with
orbifolding for $\Gamma\subset SL(2,\IR)_L \times SL(2,\IR )_R$.
Let us focus on the special case of a Hecke congruence subgroup
\eqn\hcke{
\Gamma = \Gamma_0(N) \subset SL(2,\IZ)\subset SL(2,\IR)_L
}
so we are considering the spacetime
\eqn\a{
\Gamma \backslash \widetilde{SL(2,\IR )}
}
which may be pictured as a modular curve, evolving in time.
The cusps of the modular curve trace out null lines at infinity.

Some of the on-shell scalar fields of supergravity are constructed
from $L^2(\Gamma \backslash \widetilde{SL(2,\IR )})$.  The
boundary asymptotics of these forms are, of course, well-studied in
number theory, and in this way the
the  ``scattering matrix'' for Eisenstein series \Iwaniec,
 finds an interpretation in AdS/CFT.

\newsec{Lecture II: Arithmetic and Attractors}

\subsec{Introduction}

Modular forms, congruence subgroups, elliptic curves,
are all mathematical objects of central concern both to
number theorists and to some physicists.
A nice illustration of the common interests physicists
and mathematicians share in this area is the excellent
predecessor to the present proceedings \leshouchesvol.
In this lecture, we will be discussing the
possibility that there are interesting arithmetical
issues connected with the theory of string compactification.
We will mostly be reviewing \MoorePN\MooreZU, although
we will make several new points along the way.

While there are many common tools and mathematical objects
in string compactification and in number theory,  one often finds that
the detailed
questions of the number theorists and the string
theorists are quite different.
As an illustration of this point, in   string perturbation theory we
encounter the elliptic curve
\eqn\ellcurv{
E_\tau :=  \IC/(\IZ + \tau \IZ)
}
but in string perturbation theory there isn't any compelling reason to
restrict attention to elliptic curves defined over
$\IQ$ (or any other  number field).
Moreover, one   can argue that compactification on arithmetic
varieties cannot be special. Firstly, physical quantities such as masses,
scattering amplitudes, etc. change continuously
with the moduli of compactification varieties.
Secondly, different arithmetic models for the same
variety over $\IC$ have different number-theoretic properties.
For example, the elliptic curves $y^2 = x^3 + n$ for $n\in \IZ$
are in general inequivalent over $\IQ$, although they are of course
equivalent over $\IC$.

In spite of the above discouraging remarks, in this
lecture we'll   present a little evidence for the contrary viewpoint.
We begin by describing the
``attractor mechanism.'' This is a mechanism that
distinguishes certain complex structure moduli as
being special.
The point of this talk is that the ``attractor mechanism''
for susy black holes provides a
framework which naturally isolates
certain arithmetic varieties.
At the level of slogans, one can say that
{\it supersymmetric black holes for $IIB$ string
theory on $CY$ 3-folds select arithmetic varieties. }
Whether this is really true for arbitrary Calabi-Yau 3-folds, and
whether the arithmetic of these varieties has
physical significance is still an open problem. We will
indicate some ways in which the physics and
arithmetic are related.

Some closely related works, which we will not
review here, include  \MillerAG\LynkerJN\LynkerAJ\LynkerHJ.

 \subsec{ The ``attractor equations''}

\def\cmtld{{\widetilde{\CM}}}

The ``attractor equations'' are conditions on
the Hodge structure of Calabi-Yau manifolds.
They were introduced in the context of studies of
black holes in Calabi-Yau compactification of
string theory, for reasons we will explain in the
next sections, by S. Ferrara, R. Kallosh, and A. Strominger
in \fks\StromingerKF.

Let $X$ be a compact Calabi-Yau 3-fold, and let
$\widetilde{\CM}  $ be the Teichmuller space  of complex structures on  $X$.
 Consider an integral vector $\gamma \in H^3(X,\IZ)$.
Given a complex structure $t\in \cmtld$ we have a
Hodge decomposition:
\eqn\hdgdec{
\gamma = \gamma^{3,0} + \gamma^{2,1}  + \gamma^{1,2} + \gamma^{0,3}
}

{\bf Definition}: The {\it attractor equations}
on the complex structure determined by   $\gamma$ are the equations
\eqn\aaa{
\mathboxit{
\gamma = \gamma^{3,0} +   \gamma^{0,3}}
}

Equivalently, since $h^{3,0}=1 $, we can choose a generator $\Omega$ for  $H^{3,0}(X)$
and write instead:
\eqn\othfrm{
2 {\rm Im} (\CC  \Omega) = \gamma  \in H^3(X;\IZ)
}
for some constant $\CC$. In order to make contact with the
literature let us write these equations yet another way.
 Choose a symplectic basis $\alpha^I,\beta_I$ for $H_3$.
Define ``flat coordinates'':  $X^I = \int_{\alpha^I} \Omega, F_I =\int_{\beta_I} \Omega $.
Then the attractor equations become:
\eqn\newatt{
\eqalign{
\bar \CC X^I - \CC \bar X^I & = i p^I \cr
\bar \CC F_I - \CC \bar F_I & = i q_I \cr}
}
In the remainder of the lectures we will discuss three different ways in which
these equations show up in string compactification.

\subsec{First avatar: BPS states and black holes in   IIB strings on $M_4 \times X$ }

\subsubsec{Compactification of IIB string theory on
 $M_4 \times X$ }

In order to set some notation let us consider briefly some aspects of
compactification of type IIB string theory on $M_4 \times X$,
where $M_4$ is a Lorentzian 4-manifold, such as $\IR^{1,3}$, or a spacetime
asymptotic to $\IR^{1,3}$.
If $X$ has generic $SU(3)$ holonomy then there is a unique covariantly
constant spinor, up to scale and hence the 32-real dimensional space of
supercharges is reduced to an 8-real dimensional space. That is, the low energy
supergravity has $\CN=2$ supersymmetry.

\lref\deWitPX{
B.~de Wit, P.~G.~Lauwers and A.~Van Proeyen,
``Lagrangians Of N=2 Supergravity - Matter Systems,''
Nucl.\ Phys.\ B {\bf 255}, 569 (1985).
}

\lref\AndrianopoliCM{
L.~Andrianopoli, M.~Bertolini, A.~Ceresole, R.~D'Auria, S.~Ferrara, P.~Fre and T.~Magri,
``N = 2 supergravity and N = 2 super Yang-Mills theory on general scalar  manifolds: Symplectic covariance, gaugings and the momentum map,''
J.\ Geom.\ Phys.\  {\bf 23}, 111 (1997)
[arXiv:hep-th/9605032].
}

$d=4$, $\CN=2$ supergravities are highly constrained physical systems
\deWitPX\AndrianopoliCM. For our purposes we only need to
know that there are a collection of complex scalar fields in a nonlinear
sigma model of maps $t: M_4 \to \cmtld$. (These are the ``vectormultiplet
scalars.'') In addition there is an abelian
 gauge theory with gauge algebra $u(1)^{b_3/2}$, where $b_3$ is the
Betti number of $X$. These vector fields
arise from the self-dual 5-form of IIB supergravity in 10-dimensions and
hence the theory is naturally presented without making a choice of
electric/magnetic duality frame.  The total   electric-magnetic fieldstrength:
\eqn\totalfs{
\eqalign{
\CF & \in \Omega^2(M_4;\IR) \otimes H^3(X;\IR) \cr
}
}
satisfies a self-duality constraint.
\eqn\sdconst{
\CF = *\CF
}
in ten dimensions.
The constraint \sdconst\ can be usefully expressed in
terms of the self-dual and anti-self-dual projections of the
two-form on Lorentzian spacetime as:
\eqn\a{
\CF = \CF^- + \CF^+\qquad \CF^- \in \Omega^{2,-}(M_4;\IC) \otimes
\biggl( H^{3,0}(X) \oplus H^{1,2}(X) \biggr)
}
Here we have assumed $b_1(X)=0$ for simplicity. Otherwise we need to decompose the
cohomology of $X$ into its primitive parts.

While there are many other fields in the supergravity, for our
purposes we need only
worry about the  fields described above together with the metric $g_{\mu\nu}$ on $M_4$.
These three fields are governed by the action
\eqn\bosact{
I_{\rm boson} = \int_{M_4}\sqrt{g} R + \parallel \nabla t \parallel^2 + {1 \over  8 \pi} {\rm Im}
(\CF^- , \CF^-)_{H^3}
}
where we use the natural Weil-Peterson (a.k.a. Zamolodchikov) metric on $\cmtld$
and
$(\gamma_1, \gamma_2)_{H^3}  = \int_X \gamma_1 * \gamma_2$.

\subsubsec{Superselection sectors}

Consider Hamiltonian quantization of the theory described in
the previous section, say, on $\IR^3\times time$. There will be a
Hilbert space of states decomposing into superselection sectors
described by absolutely conserved charges. The charge group is
$K^1(X)$, but for our purposes, we will focus on $H^3(X,\IZ)$.
We will
 interpret the vector $\gamma\in H^3(X,\IZ)$ in the attractor
equations as specifying a superselection sector. Semiclassically
we put a boundary condition at spatial infinity on the
electromagnetic flux:
\eqn\flxinf{
\int_{S^2_\infty} \CF = \gamma \in H^3(X,\IZ)
}
Thus, we split the   Hilbert space   into superselection sectors:
\eqn\sslc{
\CH = \oplus_{\gamma} \CH_{\gamma}
}
and interpret $\gamma$ as a vector of electric {\it and}
magnetic charges for the $\half b_3(X)$ $U(1)$ gauge fields.

The $\CN=2$ supersymmetry algebra acts on the Hilbert spaces
$\CH_{\gamma}$ and has a nonzero ``central charge''
in each of these sectors. That is, the algebra is realized as
\eqn\ntwoalg{
\{ Q_{\dot \alpha i}, Q_{\beta j} \} = \delta_{ij} \gamma^\mu_{\dot \alpha \beta} P_\mu
\qquad
\{ Q_{\alpha i} , Q_{\beta j} \} = \epsilon_{\alpha\beta}
\epsilon_{ij} Z
}
where the central charge $Z$ depends on the value of the
scalar fields $t(\infty)$ and the charge vector $\gamma$.

\bigskip
\noindent
{\bf Definition/Proposition}: For $\gamma\in H^3(X;\IZ), t(\infty)\in \widetilde{\CM} $,
  the central charge is:
\eqn\a{
Z(t;\gamma) := e^{K/2} \int \Omega\wedge \gamma  \qquad e^{-K } := i \int_X \Omega \wedge \bar \Omega > 0
}

This is a result of a direct computation when one expresses the supercharges
$Q_{\alpha i}$ in terms of the fields and computes the relevant Poisson
brackets. However, for the mathematical reader one can simply take it as
a definition of $Z(t;\gamma)$.

\subsubsec{ Attractor points minimize BPS mass}

Now we finally meet the attractor equations when we ask about properties of
``BPS states.'' Let us first explain this term. A simple consequence
of the algebra \ntwoalg\ is that in the sector $\CH_{\gamma}$
the Hamiltonian is bounded below
\eqn\bddbl{
H \geq \vert Z(t;\gamma)\vert
}

\noindent
{\bf Definition:}  A {\it BPS state} is a state $\Psi \in \CH_{\gamma}$ which saturates
the bound \bddbl.

BPS states have proven to be extremely useful in investigations of
nonperturbative physics because the associated representations of
the supersymmetry algebra have rigidity properties, and are hence
unchanged, under variation of parameters such as coupling constants.
Examples of BPS states in the present context are provided by
D3 branes wrapped on calibrated 3-cycles in $X$. The mirror of such
states are associated with certain
elements of the derived category of coherent sheaves
on the mirror of $X$.

Because of their importance we
 are interested in the behavior (and existence)  of BPS states
as a function of moduli. It is here that the attractor equations
enter the picture.
One useful diagonostic of the existence of such states is
associated with   the behavior of
 $\vert Z(t;\gamma) \vert^2 $  as a function on $ \widetilde{\CM}$.
The first key result, due to \fks\StromingerKF\fk\FerraraTW\ is

\bigskip
\bigskip
\ndt
{\bf Theorem}  If
$\vert Z(t;\gamma) \vert^2 $
has a  stationary point in $t\in \widetilde{\CM}$,
i.e., $d \vert Z(t;\gamma) \vert^2=0$, then,
\medskip
\ndt
a.) If $Z(t;\gamma) =0$, then   $\gamma \in H^{2,1}\oplus H^{1,2}$,
$t\in \CD_\gamma\in Div(\widetilde{\CM})$.

\bigskip
\ndt
b.) If $Z(t;\gamma) \not=0$,  then  $\gamma \in H^{3,0}\oplus H^{0,3}$, $t=t_*$ is an isolated minimum.

%

The proof is extremely simple, so let us include it here.
Choose $\Omega(s)$ to vary holomorphically
with $s \in \cmtld$ a local holomorphic parameter. Then, if
$\hat \gamma$ is Poincar\'e dual to $\gamma$,
\eqn\locvar{
\p_s \vert Z(\gamma) \vert^2 = \int_{\hat \gamma}\Biggl( \p_s \Omega -
{ \langle \p_s \Omega, \bar \Omega \rangle \over  \langle \Omega, \bar \Omega \rangle} \Omega\Biggr) \cdot
{\int_{\hat \gamma} \bar \Omega \over   i \int_X \Omega \wedge \bar{\Omega} }
}
Now,      $\gamma$ has a Hodge decomposition:
\eqn\a{
\gamma = \gamma^{3,0} +  \gamma^{2,1} +   \gamma^{1,2} +   \gamma^{0,3}
}
Stationarity of $\vert Z(t;\gamma)\vert^2$
 implies that   $Z=0$ or,   $Z\not=0$ and,
using $T^{1,0}\CM \cong H^{2,1}(X_3)$,    $\gamma^{2,1}=0$.
Since $\gamma$ is real this in turn implies
$\gamma=  \gamma^{3,0}  +  \gamma^{0,3}$.

In case  $(b)$ we have a local minimum. To see this we compute
\eqn\alcmin{
\eqalign{
\p_i\p_j \vert Z \vert^2 & =0 \cr
\p_i \pb_{\bar j }   \log[ \vert Z(\gamma) \vert^2 ]
& =
- \p_i \pb_{\bar j }  \log[ i \int_X \Omega \wedge \bar{\Omega} ] =
g_{i \bar j} \cr}
}
so the stationary point is a nondegenerate minimum if
the Weil-Peterson metric is nonsingular. That is, if the attractor point is at a
regular point in $\cmtld$. (We call such a point a ``regular attractor point.'')

\subsubsec{  Attractive fixed points and   Black Holes   }

Let us now consider the relation to black holes. Black holes are
certain solutions to (super-)gravity with special  causality properties
implied by a horizon.
The black holes we will consider are
``extremal.'' They have a maximal amount of allowed charge for a given
mass, and do not radiate. Semiclassically, they correspond to states
in the Hilbert space   $\CH_\gamma$ described in section 3.3.2.
Semiclassically, we describe these states as field configurations
satisfying the equations of motion of   supergravity.

We are going to focus on
static, spherically symmetric,   black holes of charge $\gamma$.
\foot{The seemingly innocent restriction to spherical symmetry
introduces important limitations, as described briefly in the next
subsection.}
Moreover, we will want to consider ``supersymmetric black holes.''
These conditions force the ansatz for the fields:
\eqn\fldans{
\eqalign{
ds^2 & = - e^{2 U(r)} dt^2 + e^{-2 U(r)} (dr^2 + r^2 d\theta^2 + r^2 \sin^2\theta d\phi^2) \cr
\vec E  & = e^{2U(r)} {\hat r \over r^2} \otimes Im(\gamma^{2,1} + \gamma^{0,3})  \cr
\vec B  & =  {\hat r \over r^2} \otimes Re(\gamma^{2,1} + \gamma^{0,3})  \cr
t^a & = t^a(r) \cr}
}
Here we have chosen a time direction and  $\vec E_i dx^i  = \CF_{0i} dt dx^i$
while $\vec B_i dx^i = *_3 \half \CF_{jk}    dx^j dx^k$.

The adjective ``supersymmetric black holes'' means  in this context
that the supersymmetric variation of the fermionic fields vanishes.
This imposes nontrivial differential equations on the bosonic fields.
The supersymmetry variations have the schematic form:
\eqn\susvar{
\eqalign{
{\rm gravitino} \qquad \delta \psi & \sim \nabla \epsilon + \Pi^{0,3}(\CF^-) \cdot \epsilon \cr
{\rm gaugino} \qquad \delta \lambda & \sim \dsl t\cdot \epsilon + \Pi^{2,1}(\CF^-) \cdot \epsilon \cr}
}
where $\epsilon$ is a spinor for the supersymmetry variation,
$\nabla$ is a spinor covariant derivative, $\dsl$ is a Dirac operator,
and $\Pi^{0,3},\Pi^{2,1}$ are the corresponding projection operators to the indicated
Hodge type.

Substitution of the
ansatz \fldans\ into the equations
$\delta \psi = \delta \lambda = 0$ yields a system of
first order ordinary differential equations in the radial variable
$r$. These equations can in turn be interpreted as defining a
dynamical system on the Teichmuller space $\cmtld$ as follows.
Let $\rho:=1/r$, and define $\mu:= e^{-U(r)}$. Then
\eqn\vgr{
\delta \psi=0 \qquad\qquad \rightarrow \qquad\qquad {d\mu\over d \rho} = \vert Z(t(r);\gamma)\vert
}
implies  $\mu$ is monotonically increasing as $r \to 0$. We can therefore use it as a
flow parameter. Now the equation
\eqn\dwlr{
\delta \lambda=0 \qquad\qquad \rightarrow \qquad\qquad
\mu {d t^a \over d\mu} = - g^{a\bar b} \p_{\bar b} \log \vert Z\vert^2
}
implies that we have gradient flow in $\cmtld$ to the minimum of $\vert Z\vert^2$.
The horizon of the black hole appears when there is a zero in the coefficient of
$g_{00}$. This happens when  $e^{2U(r)}\to 0$, hence at $\mu\to \infty$.

The attractor equations are the fixed point equations for the flow \dwlr\
\eqn\fxdpr{
t(r) \rightarrow t_*(\gamma) \qquad such \quad that\qquad \gamma =
\gamma^{3,0} + \gamma^{0,3}
}
this easily follows since
\eqn\a{\hat \gamma^{2,1}=0  \rightarrow t(r) = t_*(\gamma) }
At this fixed point
\eqn\fixu{
e^{-U_*}=1 + Z_*/r
}
where $Z_* := Z(t_*(\gamma);\gamma)$, and hence the near  horizon geometry
is $AdS_2 \times S^2$:
\eqn\a{
ds^2 = - {r^2 \over  Z_*^2} dt^2 + Z_*^2 { dr^2 \over  r^2} +
Z_*^2 (d\theta^2 + \sin^2\theta d\phi^2)
}
Note that the horizon area is
\eqn\horar{
{ {\rm Horizon \ Area }\over 4 \pi} = \vert Z(t_*(\gamma);\gamma)\vert^2 := Z_*^2
}

\subsubsec{ Summary \& Cautionary Remarks}

In summary, at the horizon of a susy black hole, the complex structure moduli
of the Calabi-Yau $X$ is fixed at an isolated point $t_*(\gamma)$
such that $\gamma = \gamma^{3,0} + \gamma^{0,3}$.
This is also the point at which the mass of states in
$\CH_{\gamma}^{BPS} $ is minimized.

A remarkable prediction of this picture, in the spirit of the
Strominger-Vafa computation is that
\eqn\a{
\log \dim \CH_\gamma^{BPS} \sim \pi \vert Z(t_*(\gamma);\gamma)\vert^2
}
for large charges $\gamma$.
\foot{Reference \MillerAG\ attempts to make this statement a little more precise.}
However, it is important to remark at this
point that we have oversimplified things somewhat.  In fact, the
dynamical system  can have several basins of attraction \MoorePN.
The multiple-basin phenomenon has been explored
in some depth in the papers of F. Denef and collaborators
\refs{\DenefNB,\DenefXN,\DenefRU,\BatesVX}.  In particular, Denef et. al.'s investigations have shown
that when enumerating BPS states, and accounting for entropy
 it is quite  important not to restrict attention to the spherically
symmetric black holes. This leads to the fascinating subject of
  ``split attractor flows,'' which clarify considerably the existence of the
multiple basins of attraction. Regrettably, all this is outside the scope of these
lectures.

\subsec{ Attractor points for $X = K3 \times T^2$  }

Now that we have described the significance of the attractor equations
for black holes and BPS states let us consider some examples of solutions
to these equations. We will focus on the elegant example of the
Calabi-Yau $K3 \times T^2$ and comment on other examples in section 3.6
below.
Let us choose $a$ and $b$ cycles on $T^2$ so that we have an isomorphism
\eqn\iskt{
H^3(K3 \times T^2,\IZ) \cong H^2(K3;\IZ) \oplus H^2(K3;\IZ)
}
Using \iskt\    can take $\gamma = p\oplus q$, with $p,q\in H^2(K3;\IZ)$:
It is easy to solve  the
equations:
\eqn\a{
\eqalign{
2 \Im \bar C \int_{a \times \gamma^I} dz \wedge \Omega^{2,0} & = p^I \cr
2 \Im \bar C \int_{b \times \gamma_I} dz \wedge \Omega^{2,0} & = q_I \cr}
}
and the answer is
\eqn\answr{
    \Omega^{3,0}
= dz \wedge (q- \bar \tau p)
}
where $dz$ is a holomorphic differential on $T^2$.
By the Torelli theorem, the
 complex structure of the $K3$ surface
is determined by $\Omega^{2,0}=( q - \bar \tau p)$.
Now, note that
\eqn\const{
\eqalign{
\int_S \Omega^{0,2} \wedge \Omega^{0,2} = 0
&
\Rightarrow p^2 \tau^2 - 2 p\cdot q \tau + q^2 =0 \Rightarrow \cr}
}
\eqn\taup{  \tau=
 \tau(p,q) := {p\cdot q +   \sqrt{D} \over  p^2 }
}
\eqn\deep{
D = D_{p,q} := (p\cdot q)^2  - p^2 q^2
}
Thus, we conclude that a regular    attractor point exists for $D_{p,q}<0$ and,
for such charge vectors
\eqn\areabh{
{A\over 4\pi} =
\vert Z_*\vert^2 =\sqrt{-D_{p,q}} =  \sqrt{p^2 q^2 - (p\cdot q)^2}
}

\subsubsec{ Attractive $K3$ Surfaces}

Let us analyze the meaning of the above attractor points more closely.
Let $S$ be a K3 surface. We may then define its
Neron-Severi lattice  $NS(S) := \ker\{ \sigma \rightarrow \int_\sigma \Omega^{2,0}\}$.
The rank of the lattice $NS(S)$ is often denoted $\rho(S)$.
We define the transcendental lattice $T_S :=  (NS(S))^\perp$.
The generic K3 surface is not algebraic and hence  $NS(S) = \{0 \} $.
for the generic algebraic $K3$,  $NS(S) = H\IZ$, and
$\rho(S) = 1$. For the generic   elliptically fibered
$K3$, $NS(S) = B\IZ \oplus F \IZ$, and hence
$\rho(S) = 2$. For the attractor points,
  $NS(S) = \langle p, q \rangle^\perp \subset H^2(K3;\IZ)$ has rank $  \rho(S) = 20$
and
\eqn\tnslst{
H^{2,0} \oplus H^{0,2} = T_S \otimes \IC
}
These $K$ surfaces  are unfortunately called ``singular K3 surfaces'' in the
literature, but they are definitely not singular. Sometimes they are
called   ``exceptional K3 surfaces.''
We will refer to them as ``attractive K3 surfaces,'' because they {\it are}
rather attractive.

Rather amusingly, from \areabh\ we see that the area of a unit
cell in $T_S$ is precisely the horizon area $A/(4\pi)$ of the
corresponding black hole!

\subsubsec{ Attractive $K3$ surfaces \& Quadratic Forms}

There is a beautiful description of the set of attractive $K3$ surfaces
in terms of binary quadratic forms. This is summarized by the theorem
of Shioda and Inose \shioda:

\bigskip
\noindent {\bf Theorem} There is
a 1-1 correspondence between
attractive K3 surfaces $S$
and  $PSL(2,\IZ)$ equivalence
classes of positive even binary quadratic forms.

In one direction the theorem is easy. Given a surface $S$ we construct the
quadratic form:
\eqn\stofrm{
T_S = \langle t_1, t_2 \rangle_{\IZ} \leftrightarrow \pmatrix{ t_1^2& t_1 \cdot t_2 \cr t_1 \cdot t_2 & t_2^2 \cr}
}

The converse is rather trickier. Given
\eqn\a{
Q = \pmatrix{ 2a & b  \cr b & 2 c \cr} \qquad a,b,c \in \IZ
}
we first consider the abelian variety $A_Q = E_{\tau_1} \times E_{\tau_2}$
where
\eqn\modprs{
\tau_1 = {-b + \sqrt{D} \over  2 a} \qquad \tau_2 = {b + \sqrt{D} \over  2 }= -c/\tau_1
}
One's first inclination is to construct the associated Kummer variety, which is
the resolution of $A_Q/\IZ_2$. Such $K3$ surfaces are indeed attractive K3 surfaces,
but do not encompass all such surfaces. Shioda and Inose introduce a clever
construction involving a pencil of elliptic curves with $E_8$ singularities to
construct a branched double cover $Y_Q$ which is itself a $K3$ surface. It is
these $Y_Q$ which account for all attractive K3 surfaces and are in 1-1 correspondence
with the quadratic forms.

Thanks to the Shioda-Inose theorem it is now trivial to describe the
attractor  points

\bigskip
\ndt
{\bf Corollary.}
Suppose  that $  \langle p, q \rangle\subset H^2(K3;\IZ)$
is a {\it primitive sublattice}. Then
the attractor variety $X_{p,q}$ determined by $\gamma= (p,q)$
is
\eqn\al{
E_{\tau(p,q)} \times Y_{2Q_{p,q}}
}
where $\tau(p,q)$ is given by
\eqn\am{
\tau(p,q) = {p\cdot q + i \sqrt{-D} \over  p^2 }
}
and $Y_{Q_{p,q}}$ is the Shioda-Inose K3 surface associated
to the even quadratic form:
\eqn\an{
2 Q_{p,q} :=
\pmatrix{ p^2 & -p\cdot q \cr - p\cdot q & q^2 \cr}
}
The variety   is a double-cover of a Kummer
surface constructed from
\eqn\ap{
X_{p,q}=
Y_{2 Q_{p,q}} \times E_\tau \rightarrow Km\Biggl( E_{\tau(p,q)} \times E_{\tau'(p,q)}\Biggr)  \times E_{\tau(p,q)}
}
with
\eqn\aq{
\tau'(p,q)= {-p\cdot q + i \sqrt{-D} \over  2 }.
}

\subsec{  $U$-duality and horizon area}

We have now described the attractor varieties. They are beautiful
and have the interesting arithmetic property that all their periods
are valued in quadratic imaginary fields. We will see in a moment that
there is much more nontrivial arithmetic associated to them. However,
we would like to know   whether this rich arithmetic structure has
any     physical significance. In this section we attempt to
make a connection to physics.

In string string theory there are ``duality groups.''
These are  arithmetic groups
which map two different charges with ``isomorphic physics.''
It is thus a natural question to ask how  $U$-duality acts on the attractor varieties.
For $IIB/K3 \times T^2$ the $U$-duality group is
\eqn\a{
U = SL(2,\IZ) \times O(22,6;\IZ)
}
The pair of
(Electric,Magnetic) charges $(p,q)$, has
$p,q\in II^{22,6}$ and forms a doublet under $SL(2,\IZ)$.
In these lectures we are suppressing certain other fields in
the supergravity, and hence we are restricting attention to
 $ p,q \in  H^2(K3,\IZ)\cong II^{19,3} \subset II^{22,6}$,
so the duality group should actually be considered to be
$SL(2,\IZ) \times O(19,3;\IZ)$.

Now, to  a charge $\gamma = (p,q)$ we associate:
\eqn\dfmtx{
2 Q_{ p,q} :=
\pmatrix{ p^2 & -p\cdot q \cr - p\cdot q & q^2 \cr}
}
This is manifestly  $T$-duality invariant while under $S$-duality
\eqn\sdtmr{
Q_{p,q} \to Q_{p',q'} = m Q_{p,q} m^{tr} \qquad \quad m\in SL(2,\IZ)
}
Note that the   near-horizon metric only
depends on the discriminant:
\eqn\a{
{ A(\gamma) \over  4 \pi} =   \sqrt{-D_{p,q}}
}
Thus, $A(\gamma)$ is invariant under
$U(\IZ)$. Still, it might be that $U$-duality-inequivalent
charges $\gamma$ have the same $A(\gamma)$. Asking this
question brings us to the topic of class numbers.

\subsubsec{ Class Numbers }

The equivalence of integral binary quadratic forms:
\eqn\qfeva{
m \pmatrix{a & b/2 \cr b/2 & c \cr} m^{tr}
= \pmatrix{a' & b'/2 \cr b'/2 & c' \cr} \qquad
  m\in SL(2,\IZ)
}
is one of the beautiful chapters of number theory.
A major result of the efforts of   Fermat, Euler, Lagrange, Legendre, and Gauss
is a deep understanding of the nature of this equivalence.
For a nice discussion of the subject see \leshouchesvol\ or
\cox. ( Reference \MoorePN\ contains   further references. ) Let us summarize a
few facts here.

 Assume, for simplicity, that the quadratic form is
primitive, that is, that
${\rm g.c.d.}(a,b,c)=1$.
There are a finite number of {\it inequivalent} classes
under $SL(2,\IZ)$.
The number of classes is the {\it class number},
denoted $h(D)$,
where
\eqn\disc{D=b^2-4 a c}
is the discriminant. We will be focussing on the case $D<0$.
It is a nontrivial fact that
one can define the structure of an abelian group
on the set of classes $C(D)$. When $D$ is a {\it fundamental
discriminant} then the   class group $C(D)$ is
isomorphic to  the group of ideal classes of the   quadratic
imaginary field
\eqn\quadfld{
K_D:= \IQ[i \sqrt{\vert D \vert} ] := \{ a + i b \sqrt{\vert D \vert} :
a,b \in \IQ \}  }
A ``fundamental discriminant'' is a $D$ such that it is the field
discriminant of a quadratic imaginary field. This turns out to mean
that $D=1 \mod 4$ and is squarefree, or, $D=0 \mod 4$, $D/4 \not=1 \mod 4$,
and $D/4$ is squarefree.

A convenient device for what follows is to associate to a  quadratic form
\eqn\qutau{
Q = \pmatrix{a & b/2 \cr b/2 & c\cr}
}
a point $\tau \in \CH$ via:
\eqn\a{a x^2 + b xy  + cy^2 = a \vert x - \tau y \vert^2
}
that is,
\eqn\gaus{
\tau = {-b + \sqrt{D} \over  2 a}
}
then
$SL(2,\IZ)$ transformations \qfeva\ act on $\tau$ by fractional linear transformations,
and hence the  inequivalent classes  may be labelled by points
$\tau_i \in \CF$:

\bigskip
\noindent
{\bf Example:} $D=-20$:
\eqn\expli{
\eqalign{
\pmatrix{1 & 0 \cr 0 & 5 \cr} \qquad
&
 x^2 + 5 y^2 \qquad\qquad  \tau_1=i\sqrt{5} \cr
 \pmatrix{2 & 1\cr 1& 3\cr} \qquad
&
 2 x^2   + 2 xy + 3 y^2 \qquad \tau_2={-1 + i\sqrt{5}\over  2}  \cr}
}
The class group is $\IZ_2$, $[\tau_1]$ is the identity element, so the
class group has multiplication law:
\eqn\clsgp{
[\tau_2]*[\tau_2] = [\tau_1].
}

\subsubsec{  $U$-Duality vs. Area (or Entropy) }

It follows immediately from the previous section that
there can be $U$-duality inequivalent BPS black holes
with the same horizon area $A$. More precisely, let
$\CB\CH(D)$ denote the number of
  U-inequivalent BPS  black  holes  with $A =  4\pi\sqrt{-D}$.
We would like to give a formula for this number.
\foot{The discussion that follows assumes that a primitive
lattice $T$ defined by $(a,b,c)$ has a unique embedding into
$II^{19,3}$. Indeed, this was blithely asserted in
\MoorePN, however further reflection shows that the statement
is less than obvious. The Nikulin embedding theory
characterizes the genus of the complementary lattice $T^\perp$
in $II^{19,3}$, and the embedding is specified by the
isomorphism class of the isomorphism of dual quotient
groups $T^*/T \to (T^\perp)^*/T^\perp$. If $T^*/T$ is
$p$-elementary then theorem 13, chapter 15 of \conway\
shows that the class of $T^\perp$ is unique. When
$T^*/T$ is not $p$-elementary there are further subtleties
associated with the spinor genus of $T^\perp$. In addition,
there can be distinct isomorphisms between the dual
quotient groups. Clearly, this aspect of the counting of
$\CB \CH(D)$ needs further thought. }
Then, if  $D$ is square-free the associated forms must
be primitive and
$\CB\CH(D)  = h(D)$. More generally, since $h(D)$
counts the primitive quadratic forms of discriminant
$D$ we have
\eqn\gnsfum{
\CB\CH(D)  = \sum_{m} h(D/m^2)
}
The sum is over $m$ such that $  D/m^2 = 0,1 \mod 4$.

Now, the  number of classes {\it grows} with $\vert D \vert$.
More precisely, it follows from work of   Landau, Siegel, and Brauer
that  $\forall \epsilon>0, \exists C(\epsilon)$ with
$
h(D) > C(\epsilon) \vert D \vert^{1/2 -\epsilon}
$
Roughly speaking, we can say that
at large entropy the number of $U$-duality
inequivalent black holes with fixed area $A$ grows
like $A$. The $U$-duality inequivalent black holes
are certainly physically inequivalent, nevertheless,
the area is a fundamental attribute and the set of
black holes with area $A$ forms a distinguished class
of solutions. It is interesting to ask if there is
 some larger ``symmetry'' which unifies these.
We will give a tentative positive answer to this question in
section 3.5.6.

\subsubsec{ Complex Multiplication  }

The attractor varieties are closely related to another
 beautiful mathematical theory, the theory of complex multiplication,
which goes back to the 19th century mathematicians Abel, Gauss,
Eisenstein, Kronecker, and Weber and continues as an active subject
of research to this day. An excellent pedagogical reference
for this material is \cox. Further references can be found in
\MoorePN.

To introduce complex multiplication let us consider the elliptic
curve $E_\tau$. This is an abelian group and we can ask about
its group of endomorphisms. Note that there
 is always a map $z \rightarrow n z$, for $n\in \IZ$, because
\eqn\simscl{
n \cdot (\IZ + \tau \IZ) \subset \IZ + \tau \IZ.
}
So $End(E_\tau)$ always trivially contains a copy of $\IZ$.
However, for special values of $\tau$, namely those for which
\eqn\spectaus{
a \tau^2 + b \tau + c =0}
for some integers   $a,b,c\in \IZ$ the
  lattice has an {\it extra ``symmetry''}, that is,
$End(E_\tau)$ is strictly larger than $\IZ$, because
\eqn\a{
\omega \cdot (\IZ + \tau \IZ) \subset \IZ + \tau \IZ\qquad\qquad
\omega={D + \sqrt{D} \over  2}
}
Here again $D=b^2-4ac$.
We say that   ``$E_\tau$ has complex multiplication by $z \rightarrow \omega z$''

To see that $E_\tau$ has wonderful properties, we choose a Weierstrass model
for $E_\tau$
\eqn\weirmod{
\eqalign{
y^2 & = 4 x^3 - c(x+1)
\qquad c = {27 j \over j- (12)^3} \qquad j\not= 0,1728\cr
y^2 & = x^3 + 1 \qquad  j=0 \cr
y^2 & = x^3 + x \qquad  j= 1728\cr}
}
and consider next some remarkable aspects of the $j$-function.

\subsubsec{ Complex multiplication and special values of
$j(\tau)$ }

The first main theorem of complex multiplication states

\bigskip
\noindent
{\bf Theorem } Suppose $\tau$ satisfies the quadratic
equation $a \tau^2 + b \tau + c=0$ with $gcd(a,b,c) = 1$,
and $D$ is a fundamental discriminant.
Then,
\bigskip
\bigskip
i.) $j(\tau)$ is an algebraic integer of degree
$h(D)$.
\bigskip
ii.) If $\tau_i$ correspond to the distinct
ideal classes in $\CO(K_D)$,
the minimal polynomial of $j(\tau_i) $ is
\eqn\a{
p(x) = \prod_{k=1}^{h(D)} (x-j(\tau_k)) \in \IZ[x]
}
Moreover: $\widehat{K_D} := K_D(j(\tau_i))$ is Galois over $K_D$ and
{\it independent of $\tau_i$} (it is a ``ring class field'' ).

Note that $\tau \to j(\tau)$ is a complicated transcendental function.
Thus, the theorem of complex multiplication is truly remarkable.

\bigskip
\noindent
{\bf Examples:}

\eqn\modrsp{
\eqalign{
\pmatrix{1 & 0 \cr 0 & 1 \cr} \qquad
j(i  )& = (12)^3 \qquad p(x)=x  - 1728 \cr
\pmatrix{1 & 0 \cr 0 & 2 \cr} \qquad
j(i \sqrt{2})& = (20)^3 \qquad p(x)=x  - 8000 \cr
\pmatrix{1 & 0 \cr 0 & 5 \cr}
\qquad \qquad
j(i \sqrt{5}) & = (50 + 26 \sqrt{5})^3 \cr
\pmatrix{2 & 1\cr 1& 3\cr} \qquad
 j({1+ i \sqrt{5} \over 2} )
& = (50 - 26 \sqrt{5})^3 \cr
p(x)   =  x^2 -  & 1264000\ x  -681472000  \cr}
}

\subsubsec{  The Attractor Varieties  are Arithmetic }

For us, the main consequence of the first main theorem of
complex multiplication is that the attractor varieties are
{\it arithmetic varieties}. That is, they are defined by
polynomial equations with algebraic numbers as coefficients.

Let us begin with the factor $E_{\tau}$ in the attractor variety.
Here  it follows from \weirmod\ and the above theorem  that
  $E_\tau$  has a model defined
over $\widehat{K_D} = K_D(j(\tau_i))$.

Now, let us turn to the $K3$ surface factor. The Shioda-Inose
construction begins with the abelian surface $E_{\tau_1} \times E_{\tau_2}$
defined by \modprs. Now, $j(\tau_i/c)$ is arithmetic and hence
the abelian surface is arithmetic. Moreover, forming
 the Kummer surface and   taking the branched cover
can all be done algebraically, but involves the
coordinates of the torsion points   of $E_\tau$.
Now we need the second theorem of complex multiplication:

\bigskip
\noindent
{\bf Theorem}  Let $c=27j/(j-1728)$
\eqn\a{
\eqalign{
E_\tau=\{ z: z \sim z+
&
\omega, z \sim z+\omega \tau \} \cr
& \cong \{ (x,y): y^2 = 4 x^3 - c(x+1) \} \cr}
}
The torsion points $(x,y)_{a,b,N}$ corresponding
to $z= {a + b \tau \over  N} \omega$ are arithmetic
and generate   finite abelian extensions of
$\hat K_D$. Moreover
\eqn\a{
\hat K_{N,D} = K_D(j, x_{a,b,N})
}
are ``ring class fields.''

Thus, the Shioda-Inose surface is an arithmetic surface and
we arrive at the important conclusion:
{\it The $K3 \times T^2$ attractor variety,
$Y_{2 Q_{p,q}} \times E_{\tau_{p,q}}$ is
{\it arithmetic}, and is defined over a
finite extension of $\widehat{K_D}$.}
It would actually be useful to know more precisely which
extensions the variety is defined over. This is an open
problem (probably not too difficult).

\subsubsec{ ${\rm Gal}(\bar{\IQ} /\IQ)$ action on
the  attractors  }

In the previous section we have seen that the   attractor
varieties are defined over  finite extensions of $\hat K_D$.
Therefore, ${\rm Gal}(\bar{\IQ} /\IQ)$ acts on the complex structure
moduli of attractors. What can we say about
this orbit?

Here again we can use a   result of ``class field theory'':
$\widehat{K _D}$ is Galois over $K_D $,
and $Gal(\widehat{K _D}/K_D)$ is in
fact isomorphic to the class group $C(D)$. Indeed, the isomorphism
$[\tau]   \rightarrow \sigma_{[\tau]}\in Gal(\widehat{K _D}/K_D)$
satisfies the beautiful property that
\eqn\clssiso{
\eqalign{
[\tau] & \rightarrow \sigma_{[\tau]}\in Gal(\widehat{K _D}/K_D)\cr}
}
is defined by
\eqn\sfglse{
j([\bar \tau_i] * [\tau_j]) = \sigma_{[\tau_i ] }(j[\tau_j])  }

\bigskip
\ndt
{\bf Example:} Once again, let us examine our simple example of
 $D=-20$. Here $K_D = \IQ(\sqrt{-5})$, and as we have seen
\eqn\xpdlin{
\eqalign{
D=-20  \quad & \hat K_{D=-20} =K_{-20}(\sqrt{5}) = \IQ(\sqrt{-1},\sqrt{-5}) \cr
 \langle \sigma \rangle & = {\rm Gal}(\widehat{K _D}/K_D) \cong \IZ/2\IZ\cr}
}
In this case, \clssiso\ is verified by:
\eqn\clsidn{
\eqalign{
 & (50 - 26 \sqrt{5})^3=  j({1+ i \sqrt{5} \over 2} )= j([\tau_2]*[\tau_1])   \cr
 &  =\sigma_{[\tau_2]}(j([\tau_1])) = \sigma_{[\tau_2]}\bigl[ j(i \sqrt{5})\bigr] = \sigma_{[\tau_2]} ( (50 + 26 \sqrt{5})^3)  \cr}
}

Now, since   ${\rm Gal}(\bar{\IQ}/\IQ)$  permutes the
different $j(\tau_i)$ invariants it   extends the $U$-duality
group and ``unifies'' the different attractor points at
discriminant $D$. In this sense, it answers the question
posed at the end of section 3.5.2.  Because we have not been
very precise about the field of definition of the attractor
varieties we cannot be more precise about the full Galois
orbit. This, again, is an interesting open problem.

\subsubsec{ But, the Galois group ${\rm Gal}(\bar {\IQ}/{\IQ})$ is not a symmetry of the BPS  mass spectrum}

The physical role (if any) of the Galois group action mentioned above
remains to be clarified. We would like to stress one important point:
The BPS mass spectrum at different attractor points related by the Galois
group action are in general {\it different}, so the   Galois action is
not a symmetry in any ordinary sense.

A simple example of this
is illustrated by the Calabi-Yau manifold   $X=(S\times E)/\IZ_2$,
where $S$ is the double cover of an Enrqiques surface.
The BPS mass spectrum at
an attractor point determined by $p_0,q_0 \in II^{2,10}$
and turns out to be
\eqn\fhsv{
 \vert Z(t_*(p_0,q_0);p,q)\vert^2
=  {1\over 2 \vert D_{p_0,q_0}\vert^{3/2}}  \vert A - \tau(p_0,q_0) B \vert^2
}
$A,B$ are integers depending on $p,q,p_0,q_0$.
 Thus the   BPS mass spectrum at the
attractor point for $\gamma = p_0\oplus q_0$ is
completely determined by  the norms of
ideals in the ideal class corresponding to
$Q_{p_0,q_0}$. At inequivalent $\tau_i$  the spectra
are in general different.

There have been other
attempts at finding a physical role for the Galois group in the present
context. Some attempts
involve the action on locations of D-branes \MoorePN\GukovNW, and
there are others \LynkerJN\LynkerAJ.
In a lecture at this workshop A. Connes made a very
interesting suggestion of a relation of our discussion to
his work with J.-B. Bost on arithmetic spontaneous symmetry
breaking \connes. In this view the Galois group is a symmetry,
but the symmetry is broken.

\subsec{Attractor Points for  Other Calabi-Yau Varieties}

Let us briefly survey a few known results about attractor points
for other Calabi-Yau varieties.

\subsubsec{ $T^6$}

The story here is similar to the case of $K3 \times T^2$.
For  $IIB/T^6$ the
$U$-duality group is
$E_{7,7}(\IZ)$ \HullYS.
The charge lattice is
a module for $E_{7,7}(\IZ)$ of rank 56.
The area of the black hole horizon is  $A/4\pi = \sqrt{-D(\gamma)}$, with
$D(\gamma) = - I_4(\gamma)$, where $I_4(\gamma)$ is
Cartan's quartic invariant defining $E_7\subset Sp(56)$ \KalloshUY.

If we choose
\eqn\gvs{\gamma\in H^3(T^6;\IZ)\subset
\IZ^{56}}
then  an  explicit computation, described in \MoorePN\
shows that  the attractor variety $\IC^3/(\IZ^3 + \tau \IZ^3)$
is isogenous to $E_{\tau_0}  \times E_{\tau_0}\times E_{\tau_0}$,
where $\tau_0 = i \sqrt{I_4(\gamma)}$,
and is therefore defined over a finite   extension
of $\IQ[i \sqrt{I_4}]$.

\subsubsec{ Other  Exact CY Attractors }

Some examples of other exactly known attractors are

\item{1.} Orbifolds  of $T^6$ and  of $K3 \times T^2$.

\item{2.} The mirror of the
Fermat point $2x_0^3 + x_1^6 + x_2^6 + x_3^6 + x_4^6=0$.

\item{3.} Consider the Calabi-Yau subvariety in   $P^{1,1,2,2,2}[8]$
defined by
\eqn\twoparone{
x_1^8 + x_2^8 + x_3^4 + x_4^4 + x_5^4
- 8 \psi x_1 x_2 x_3 x_4 x_5 - 2 \tilde\phi x_1^4 x_2^4 = 0
}
{}From the formulae of Candelas et. al., in ref. \CandelasDM\  we can  find exact
attractors for $\psi=0$, via the change of variables:
\eqn\twopartwo{
\eqalign{
\tilde\phi^{-2}   = {16 z (1-z) \over  (1+ 4z - 4 z^2)^2}
\qquad &\qquad
z = -{\vartheta_2^4(\tau) \over  \vartheta_4^4(\tau) }\cr}
}
The attractor points correspond to  $\tau  = a + b i \in \IQ[i] $,
$-1<a<1, b>0$.
In fact, the last two examples are  $K3 \times T^2$ orbifolds,
as was pointed out to me by E. Diaconescu and B. Florea.

\item{4.} Any  rigid  Calabi-Yau manifold is automatically an attractor
variety. We will return to this in remark 5 in the next subsection.

\subsubsec{ Attractor Conjectures \&  Remarks}

We will now state some conjectures. It is useful
to draw the following distinction between attractor points.
The attractor equation says that there is an integral vector
\eqn\rki{
\gamma \in H^{3,0} \oplus H^{0,3}
}
It can happen that there is
a rank 2 submodule $T_X \subset H^3(X;\IZ)$ with
\eqn\rkii{   H^{3,0} \oplus H^{0,3} = T_X \otimes \IC}
We call such a point an ``attractor of rank 2.''
It is simultaneously an attractor point for
two   charges $\gamma_1,\gamma_2$ with
$\langle \gamma_1, \gamma_2 \rangle \not=0$.
If it is not of rank two we call it an
``attractor of rank 1.''

Based on the above examples one may jump to a
rather optimistic conjecture which we call the
{\it Strong Attractor Conjecture}:
Suppose $\gamma $ determines
an attractor point
$t_*(\gamma)\in \cmtld$.
Then the flat coordinates of special geometry
are valued in a number field $K_\gamma$, and $X_\gamma$ is
an arithmetic variety over some finite extension of
$K_\gamma$.
A more modest conjecture, the {\it Weak Attractor Conjecture} only asserts this for rank $2$
attractor points.

Unfortunately, there has been very little progress on these conjectures
since they were suggested in \MoorePN\MooreZU. Some salient points are the
following:

\item{1.}
All known exact attractor points are of rank two. Moreover,
 the evidence is also consistent with the conjecture that
all rank 2 attractors are orbifolds of $T^6$ and $K3 \times T^2$.
Since rigid Calabi-Yau
manifolds are necessarily rank 2 attractors, this suggestion
can perhaps be falsified by the interesting examples
mentioned in \yui. \foot{In some unpublished work, R. Bell has checked
that some of these examples are indeed arithmetic.}

\item{2.}  In the course of some
discussions with E. Diaconescu and M. Nori, Nori
was able to demonstrate that the   Hodge conjecture implies that
rank 2 attractors are indeed arithmetic. (Thus, one way
to falsify the Hodge conjecture is to produce an
example of a nonarithmetic rank two attractor.)

\item{3.} Attractor points of rank one are expected to be dense.
The density can be proved in the limit of large complex structure \MoorePN.
On the other hand, attractor points of rank two are expected to
be rare. Indeed, this issue can be addressed in a quantitative
way using computers. Sadly, a search of some $50,000$ attractor points
in the moduli space of the mirror of the quintic, performed by
F. Denef, revealed {\it no }  convincing candidates for rank two
attractors.
\foot{Briefly, Denef's method is the following. Given a
complex structure, $Re(\Omega)$ and  $Im(\Omega)$ determine a
real two-dimensional vector space $V\subset H^4(X,\IR)$. Given a
charge $Q$, Denef computes the attractor point numerically
to high precision. Now,  $Q$ is an integral vector in $V$.
Denef then constructs an orthogonal vector $P$ in $V$ using a
Euclidean metric on $H^4(X,\IZ)$. If the components of $P$
are rational then the complex structure point is a rank 2
attractor. Using the numerical value of the periods
he examines the components of $P$ and searches for
rational $P$'s using a continued fraction algorithm.
(Thus, long continued fractions are considered irrational.)
His computer then scans through a list of charges $Q$. }

\item{4.} On the positive side, we can say that should the
attractor conjectures turn out to be true they might
 imply   remarkable identities
on trilogarithms and generalized hypergeometric functions.
For an explanation of this, see section 9.3 of \MoorePN.

\item{5.} Finally, we would like to note that
there is a notion of ``modular Calabi-Yau variety'' generalizing
the notion of modular elliptic curve. The modular K3-surfaces over
$\IQ$ turn out to be attractor varieties. For a discussion of
this see \yui. The known examples of modular Calabi-Yau varieties
are rigid, and hence, automatically, are attractors.
It would be quite fascininating, to put it mildly,
if   a relationship between attractors and modular Calabi-Yau
varieties persisted in dimension 3.

\subsec{Second avatar:   RCFT   and F-Theory  }

A second, very different, way attractive K3 surfaces are distinguished in
physics is in the context of {\it F-theory}. We will now indicate how
it is that
{\it the compactification of the heterotic string
to 8 dimensions on   rational conformal field theories (RCFT's)  are dual to the F-theory
compactifications on attractive K3-surfaces.}

Recall the basic elements of $F$-theory/Heterotic duality:
\foot{For more details see \VafaXN\AspinwallMN
\ClingherUI
.}
The heterotic string on a torus $T^2$ is dual to a $IIB$ $F$-theory
compactification on a K3 surface $S$. If we fix a
hyperbolic plane: $\langle e, e^* \rangle\subset H^2(S;\IZ)$, then
$\langle e, e^* \rangle^\perp \cong II^{2,18}$, and this lattice is
identified with the charge lattice in the Narain compactification
of $F$-theory. The moduli space
 $Gr_+(2,II^{2,18}\otimes \IR) $  is interpreted in two ways.
In IIB theory it is the space of  positive definite planes  $\Pi\subset
II^{2,18}\otimes \IR$, spanned by  $Re(\Omega)$ and $Im(\Omega)$,
  which defines the complex structure of an elliptically fibered polarized K3-surface.
In the heterotic theory is it the moduli space of Narain compactifications.

\subsubsec{RCFT's for the heterotic string}

In the heterotic theory, the condition that the right-moving lattice
is generated over $\IQ$ (which corresponds to the $K3$ surface $S$ being attractive)
turns out to be equivalent to the condition that the compactification on $T^2$
is along a   {\it rational conformal field theory}. One can go further,
as shown in \MoorePN, section 10.3. Choosing decompactifications of the heterotic string
to $9$ and $10$ dimensions is equivalent to choosing a realization of the lattice
\eqn\a{
\langle w_1, w_1^* \rangle \oplus
\langle w_2, w_2^* \rangle\oplus
(E_8(-1))^2 \cong II^{2,18}
}
where $\langle w_i, w_i^*\rangle $ are hyperbolic planes. Using this decomposition the
moduli space can be realized as a tube domain in $18$-dimensional complex
Lorentzian space:
\eqn\tube{
Gr_+(2, II^{2,18}\otimes \IR) \cong \IR^{1,17} + i C_+
=\{ y = ( T, U , \vec A ) \}
}
where $C_+$ is the forward lightcone in $\IR^{1,17}$,
$U$ is the complex structure of $T^2$, $T$ is the Kahler structure, and $\vec A$ encode the
holonomy of flat $E_8 \times E_8$ gauge fields.
Under the isomorphism \tube we identify
\eqn\tubdom{
\Omega = y + w_1 - \half y^2 w_1^*
}
The conditions for a rational
conformal field theory imply that the  heterotic theory is compactified on an elliptic curve of CM
type with $(T,\vec A) $ in the quadratic imaginary field defined by $U$. Indeed, the curve has
complex multiplication by a rational integral multiple of $\bar T$.

There are further interesting relations under this duality, including relations
between the     Mordell-Weil group of the attractive elliptic K3 surface and the
 enhanced chiral algebra of the heterotic RCFT. This essentially follows from the
fact that the projection of   $p\in II^{2,18}$ onto the positive definite space:
\eqn\a{
p_R = e^{K/2} \int_p \Omega^{2,0}
}
in $F$-theory corresponds to ``right-moving momentum'' in
Narain compactification.

The above duality realizes in  part   an old dream of Friedan \& Shenker. Their idea was
to approximate superconformal field theories on Calabi-Yau manifolds by rational conformal
field theories.   Generalizations of the relation between complex multiplication and
rational conformal field theories on tori have been studied by
  K. Wendland in \wendlandthesis\WendlandMA.
A rather different relation between rational conformal field theories and complex multiplication
has been suggested by S. Gukov and C. Vafa \GukovNW. These last authors conjecture that
the superconformal field theory with target space given by a K3 surface with complex
multiplication will itself be rational.

Finally, we would like to mention the very elegant result of
S. Hosono, B. Lian, K. Oguiso, and  S.-T. Yau
in \HosonoYB, which may be phrased, roughly, as follows. Consider the map from
moduli $(T,U,\vec A=0)$ to the quadratic form characterizing the
attractor point.
The moduli $T,U$ are valued in $\IQ(\sqrt{D})$ and may therefore also be associated to
quadratic forms. Reference \HosonoYB\ shows
that the three quadratic forms are related by the Gauss product, and uses this to
give a classification of $c=2$ toroidal RCFT's.

Here is an (over)simplified version of the discussion in \HosonoYB.
When $\vec A=0$ we have
\eqn\tubdomi{
\Omega =  w_1 - TU w_1^*+ T w_2 + U w_2^*
}
A basis (over $\IR$) for the plane $\Pi$   is given by
\def\bu{\bar U}
\def\bt{\bar T}
\eqn\pibasis{
\eqalign{
\nu_1 & = w_1 + U\bar U {T-\bt \over U - \bu} w_1^* + {U\bt -\bu T \over U - \bu} w_2 \cr
\nu_2 & = {\overline{TU}- TU \over U - \bu} w_1^* + {T - \bt \over U - \bu } w_2 + w_2^* \cr}
}
while   the orthogonal plane $\Pi^\perp$ in $II^{2,2}\otimes \IR$
is spanned (over $\IR$) by
\eqn\piorthbas{
\eqalign{
\gamma_1 & = w_1 -  U\bar U {T-\bt \over U - \bu} w_1^* -{\overline{TU}- TU \over U - \bu}w_2\cr
\gamma_2 & =  - {U\bt -\bu T \over U - \bu} w_1^* -{T - \bt \over U - \bu } w_2 + w_2^*\cr}
}
Note that these  are rational vectors
iff $U,T \in \IQ[\sqrt{D}]$. In the latter case, by $SL(2,\IZ)$ transformations
we can bring them to the ``concordant'' form
\foot{For concordant quadratic forms we further require $a\vert c$, but we do not use this
condition in our discussion in sec. 3.8 below.}
\eqn\concord{
\eqalign{
U & = {b+ \sqrt{D} \over 2a} \cr
T & = {b+ \sqrt{D} \over 2a'} ={a\over a'} U \cr}
}
in which case the basis vectors simplify to
\eqn\cordo{
\eqalign{
\nu_1 & = w_1 + {c\over a'} w_1^* \cr
\nu_2 & = - {b\over a'} w_1^* + {a\over a'} w_2 + w_2^* \cr
\gamma_1 & = w_1 - {c\over a'} w_1^* + {b\over a'} w_2 \cr
\gamma_2 &  = -{a\over a'} w_2 + w_2^* \cr}
}
A straightforward computation shows that
\eqn\grtx{
(\nu_i \cdot \nu_j) ={1\over a'}  \pmatrix{2c & -b \cr-b& 2a\cr}
}
\eqn\grtxi{
(\gamma_i \cdot \gamma_j) =-{1\over a'}  \pmatrix{2c & -b \cr-b& 2a\cr}
}
If $T,U$ are associated with quadratic forms $(a,b,c)$ and $(a',b,c')$ then
$t_1 = a' \nu_1, t_2 = a' \nu_2$ is an integral basis for $\Pi$, and from \grtx\ we
see that the quadratic form of this basis is the Gauss product of the
quadratic forms associated to $T,U$.

\subsubsec{ Arithmetic properties of the K3 mirror map }

The above relation of heterotic RCFT and attractive K3 surfaces raises
interesting questions about the arithmetic properties of mirror maps.
Recall that the $j$ function itself can be viewed as a mirror map for
1-dimensional Calabi-Yau manifolds. It is natural to ask if the mirror
maps of higher dimensional Calabi-Yau manifolds have arithmetical
significance, perhaps playing the role of the transcendental functions
sought for in Hilbert's 12$^{th}$ problem.

The next case to look at is 2-dimensions. In \LianZV\ Lian and Yau studied
the mirror map for
pencils of K3 surfaces and found, remarkably, the occurance of
Thompson series. Hence the mirror map again has arithmetical properties.
The perspective on F-theory we have discussed suggests a generalization.
We may think of   F-theory compactifications in terms of a Weierstrass model:
\bigskip
\ndt
\eqn\a{
\eqalign{
Z Y^2 = & 4X^3 - f_8(s,t) X Z^2 - f_{12}(s,t) Z^3 \cr
f_8(s,t) = & \alpha_{-4} s^8 + \cdots + \alpha_{+4} t^8 \cr
f_{12}(s,t) = & \beta_{-6} s^{12}+ \cdots + \beta_{+6} t^{12}\cr}
}
In this description the moduli space is:
\eqn\a{
\eqalign{
\CM_{\rm algebraic} & = \biggl[ \{ (\vec \alpha, \vec \beta)\} - \CD\biggr]/GL(2,\IC) \cr
}
}
where $\CD$ is the discriminant variety and the action of $GL(2,\IC)$ is induced by
the action on $s,t$.   The map  $\Phi_F: y \rightarrow (\vec \alpha, \vec \beta)$, is a map from
flat coordinates to algebraic coordinates and in this sense it can be thought of as the
mirror map.  From the Shioda-Inose theorem and the theory of  complex multiplication
it is therefore natural to conjecture that
{\it The  map $\Phi_F$    behaves
analogously to the elliptic functions in the theory of
complex multiplication, i.e.,  $y^i\in K_D \rightarrow \alpha_i, \beta_i \in \hat K $
for some algebraic number field $\hat K$.  }

In \MoorePN\ some nontrivial checks on this conjecture were performed. The most
comprehensive check is to consider the map $\Phi_F$ in the limit of stable
degenerations ($T\to \infty$ in terms of the variables defined in \tube.)
In that case, one may use the results of Friedman, Morgan, and Witten
\FriedmanYQ\FriedmanSI\  to verify the statement.

\subsec{Third avatar:    Flux compactifications}

\lref\GiddingsYU{
S.~B.~Giddings, S.~Kachru and J.~Polchinski,
``Hierarchies from fluxes in string compactifications,''
Phys.\ Rev.\ D {\bf 66}, 106006 (2002)
[arXiv:hep-th/0105097].
}

\lref\KachruAW{
S.~Kachru, R.~Kallosh, A.~Linde and S.~P.~Trivedi,
``De Sitter vacua in string theory,''
Phys.\ Rev.\ D {\bf 68}, 046005 (2003)
[arXiv:hep-th/0301240].
}
\lref\BeckerGJ{
K.~Becker and M.~Becker,
``M-Theory on Eight-Manifolds,''
Nucl.\ Phys.\ B {\bf 477}, 155 (1996)
[arXiv:hep-th/9605053].
}
\lref\GranaJJ{
M.~Grana and J.~Polchinski,
``Supersymmetric three-form flux perturbations on AdS(5),''
Phys.\ Rev.\ D {\bf 63}, 026001 (2001)
[arXiv:hep-th/0009211].
}

There is a {\it third} manifestation of the attractor varieties.
It is related to a topic of current interest in string compactification,
namely, compactification with fluxes. The literature on this subject
is somewhat vast. See, for examples, \GiddingsYU\KachruAW\AshokGK\ for
some recent papers with many references
to other literature.  It turns out that this subject is closely related to
the attractor problem for Calabi-Yau {\it four-folds}.

We begin by considering  compactification of type IIB string theory on a Calabi-Yau
manifold $X_3$, now adding ``fluxes'' instead of wrapped branes, as we have
been discussing thus far. In particular, if one considers the RR and NSNS
3-forms $F_{}$ and $H_{}$, then they must be closed, by the Bianchi
identity, and they must  satisfy a quantization
condition on their cohomology classes: $[F_{}],[H_{}] \in H^3(X_3,\IZ)$.
In backgrounds with such fluxes the low energy supergravity
develops a superpotential \GukovYA, and analysis of this
superpotential shows that the supersymmetric minima with
zero cosmological constant are characterized
by complex structure and complex dilaton such that
\eqn\flxcpct{
G_{IIB}:= [F_{}] - \phi [H_{}] \in H^{2,1}_{\rm primitive}
}
for integral vectors $F_{}, H_{}$, where $\phi$ is the
axiodil (a.k.a. complex dilaton). (This can also be shown by
studying supersymmetry transformations \GranaJJ\
or by using the result of \BeckerGJ\ applied to M-theory on
$X\times T^2$.) Fluxes with
\eqn\flxcpctg{
G_{IIB}:= [F_{}] - \phi [H_{}] \in H^{2,1}_{\rm primitive} \oplus H^{0,3}
}
can also in principle be used to obtain supersymmetric AdS  compactifications with
negative cosmological constant.
\foot{In our discussion we are suppressing some important physical points.
Foremost amongst these is the fact that we need to consider an orientifold
of the compactification described above in order to have $d=4, \CN=1$ supersymmetry.
The examples below can be
orientifolded. }

Equation \flxcpct\ is usually regarded as an equation
on the complex structure of $X_3$ and the complex dilaton $\phi$.
For some classes of flux vectors $F_{}$ and $H_{}$ the solutions
are isolated points in moduli space.
\foot{There are also fluxes for which there are no solutions, and
fluxes for which there are continuous families of solutions.
A general class of examples of the latter type arise by
embedding $X$ in some ambient variety $\iota: X \hookrightarrow W$
and choosing $F_{}$ and $H_{}$ to be classes pulled back from $W$. }
Thus, \flxcpct\  is reminiscent of the attractor equations
(as noted in \CurioAE\CurioSC). However,
despite its similarity to the attractor equations,
 the condition
\flxcpct\ is in fact a very different kind of constraint on the
Hodge structure of the Calabi-Yau manifold,
since the left-hand side of \flxcpct\ is complex and nonintegral.

Despite these distinctions the flux compactification problem
is in fact related to the attractor problem, but for Calabi-Yau
{\it four-folds} $X_4$. Consider a Calabi-Yau 4-fold with
$\gamma\in H^4(X_4,\IZ)$. In analogy to section 3.3.3 above
we seek to stationarize the normalized period:
\eqn\cnt{
\vert Z(\gamma)\vert^2 = {\vert  \gamma \cdot \Omega \vert^2\over \Omega\cdot \bar \Omega } .
}
By exactly the same argument as in section 3.3.3 a stationary point
 is either a divisor where $Z(\gamma)=0$ or, if $Z(\gamma)\not=0$, a point where
$\gamma^{1,3}=\gamma^{3,1}=0$. An important distinction from the
3-fold case is that the Hessian at a critical point is not necessarily
positive definite: The first line of \alcmin\ can be nonzero since
$\gamma$ can have a $(2,2)$ component which overlaps with the second
derivatives of $\Omega$.

In the physical interpretation of the 4-fold attractor problem we
may identify   $\gamma = [G]$   as
the  cohomology class of the $G$-flux of  $M$-theory.  These compactifications
can be related to those defined by \flxcpct\ in the case where  $X_4$ is
elliptically fibered, for then we may consider an associated
$F$-theory compactification. In general, this requires the insertion of
$7$-branes in the base of the fibration, but when these coincide we can
obtain the orientifold compactifications discussed above  \SenBP.
To specialize further,   suppose $X_4 = X_3 \times T^2$. Then
$G = H_{} d \sigma^1 + F_{ } d \sigma^2$, with complex structure
$dz = d \sigma_1 + \phi d \sigma_2$ on $T^2$. Then
\eqn\gem{
G = {1\over \phi-\bar \phi} \bigl( (F-\bar\phi H) d z - (F-\phi H) d \bar z\bigr)
= {1\over \phi-\bar \phi} \bigl( G_{IIB}^* d z - G_{IIB} d \bar z\bigr)
}
so, in particular:
\eqn\gemp{
\eqalign{
G^{1,3}& = {1\over \phi-\bar \phi} \bigl( (F-\bar\phi H)^{0,3} d z - (F-\phi H)^{1,2} d \bar z\bigr)\cr
G^{0,4}& = - {1\over \phi-\bar \phi}(F-\phi H)^{0,3} d \bar z \cr}
}
and hence stationary points with $G^{1,3} = G^{0,4}=0$ correspond to supersymmetric
Minkowskian compactifications while those with $G^{0,4}\not= 0$ are related  to more
general $AdS$ compactifications.

What can we say about exact solutions to the flux compactification problem?
One  remark is that any attractor point of rank $2$ automatically
gives a solution to \flxcpctg, for some fluxes. After all, we can choose
$[F],[H]$ in the lattice $T_{X_3}$ in \rkii\ and then   choose $\phi$ so that
$[F-\phi H] \in H^{0,3}(X_3)$. Thus, all our rank two attractor
examples can be re-interpreted as flux compactifications.
For example, using \answr\taup\deep\
we could take (an orientifold of) $X_3= K3 \times T^2$ and
\eqn\simple{
\eqalign{
F & = p^2 dx\wedge q + 2 p\cdot q dy \wedge q - q^2 dy \wedge p \cr
H & = dy \wedge q + dx \wedge p \cr}
}
with $\phi = p^2 \tau$. Similarly, the example
\twoparone\twopartwo\ above provides a simple exact infinite family
with $\phi=i$. For any rational numbers $a,b$, $-1< a< 1, b>0$ we have,
from   section 8.3.2 of \MoorePN,
\eqn\simplealso{
\eqalign{
\Omega_{a,b} & :=  \gamma_1 + i  \gamma_2 \cr
 \gamma_1 & = 2  \alpha_0 -  \alpha_1 + (a+1)  \alpha_2 - (a +
b-2)  \beta^0 - 2(b+1)  \beta^1 - 4  \beta^2 \cr
 \gamma_2 & =  \alpha_1 + (b-1)  \alpha_2 - (b-a)  \beta^0 -
2(1-a)  \beta^1 \cr}
}
Here $ \alpha^i,  \beta_i$ is an integral symplectic basis. Thus,
suitable integral multiples of $ \gamma_i$ will produce examples.
For another
recent discussion of exact examples see \GiryavetsVD.

In a recent paper,   Tripathy and Trivedi
analyzed the conditions \flxcpct\ for the case when the
Calabi-Yau is $T^6$ or $K3 \times T^2$ \TripathyQW.
Their discussion can be interpreted as follows:
when the fluxes are such that the solutions admit
isolated supersymmetric vacua in complex structure
moduli space, those vacua turn out to be precisely
attractor points!

\def\a{\alpha}
\def\b{\beta}

With the benefit of hindsight we can easily describe all the
solutions in \TripathyQW\  in terms of attractor points
on $S\times T^2$ with $S$ a $K3$ surface. Choosing
a basis $dx,dy$ of 1-forms on $T^2$ we decompose
$F= \a_x dx + \a_y dy, H = \b_x dx + \b_y dy$,
where $\a_x,\a_y,\b_x,\b_y\in \Lambda\cong II^{3,19}$.
The condition  \flxcpct\ in this case
can be equivalently  written in terms of the projection of
these vectors into the plane
\eqn\cplxstr{
\Pi = \langle Re\Omega, Im\Omega\rangle \subset \Lambda \otimes \IR
}
and its orthogonal complement $\Pi^\perp \subset  \Lambda\otimes \IR$.
The condition \flxcpct\ is equivalent to the following six equations
for the projection of the vectors into $\Pi$ and $\Pi^\perp$:

\eqn\solv{
\b_x^\Pi = {1\over (\tau-\bar \tau)(\phi-\bar\phi)} (\xi \Omega + \bar
\xi \bar \Omega)
}
\eqn\solvi{
\a_x^\Pi = {1\over (\tau-\bar \tau)(\phi-\bar\phi)} (\bar \phi\xi
\Omega + \phi\bar \xi \bar \Omega)
}
\eqn\solvii{
\b_y^\Pi = {1\over (\tau-\bar \tau)(\phi-\bar\phi)} (\xi\bar\tau \Omega
+ \bar \xi \tau \bar \Omega)
}
\eqn\solviii{
\a_y^\Pi = {1\over (\tau-\bar \tau)(\phi-\bar\phi)} (\bar
\phi\xi\bar\tau \Omega + \phi\bar \xi\tau \bar \Omega)
}
\eqn\solviv{
\a_y^\perp = {(\phi\bar \tau - \bar\phi
\tau)\over  (\phi-\bar\phi)}  \a_x^\perp
+{ \phi \bar\phi (\tau - \bar \tau)\over  (\phi-\bar\phi)} \b_x^\perp
}
\eqn\solvv{
\b_y^\perp =- {( \tau - \bar  \tau)\over  (\phi-\bar\phi)}\a_x^\perp
+    {(\phi \tau - \bar \phi\bar\tau)\over  (\phi-\bar\phi)} \b_x^\perp
}
Here $\xi$ is a complex number, and $\tau$ is the period of $T^2$.
Note that $\a_x^\perp,\b_x^\perp$ are unconstrained, except that
the class $G$ is primitive iff $\a_x^\perp,\b_x^\perp$ are
orthogonal to the Kahler class $J$. We will assume the class $J$ is
rational and hence the $K3$ surface is algebraic.

When expressed this way it is manifest that for any attractor point
there is an infinite set of flux vectors associated to that point.
For, if $Y_Q$ is an attractive K3 surface associated to $(a,b,c)$ then
$\Pi$ is rationally generated. Indeed, we may take $\Omega = t_2 - \omega t_1$
where $t_1,t_2$ is an oriented basis for $\Pi$ and $\omega = (b + \sqrt{D})/2a$.
If $\tau, \phi, \xi \in \IQ(\sqrt{D})$, then all the vectors in
\solv,\solvi,\solvii,\solviii,\solviv,\solvv\ are rational.
The condition that
$\a_x^\Pi + \a_x^\perp$, etc.   lie in $\Lambda$ reduces to simple Diophantine
conditions on $\xi, \a_x^\perp,\b_x^\perp$ with infinitely many solutions.
A similar set of equations can be used to give the general solution
to \flxcpctg. In these more general solutions $\phi,\tau$ and the
attractor points can be associated with two distinct quadratic fields.

\def\tl{\tilde }
An even more explicit family of flux vacua can be obtained by
combining the 4-fold viewpoint with the
formulae \pibasis\ - \grtxi\ above. This family can be applied to
the 4-folds of the type $S\times \tl S$ where the surfaces
$S,\tl S$ can be taken to be $T^4$ or $K3$.
Denote by $T,U$ the moduli for the first factor, and by $\tl T, \tl U$
the moduli of the second factor. Similarly, a $\tilde{} $ denotes a quantity
associated with the second factor.
Choose $2\times 2$ real matrices $X,Y$ and write
\eqn\gmflx{
G = \pmatrix{\nu_1 & \nu_2} X \pmatrix{\tl \nu_1 \cr \tl \nu_2 \cr}  +
\pmatrix{\gamma_1 & \gamma_2} Y \pmatrix{\tl \gamma_1 \cr \tl \gamma_2 \cr}
}
This is automatically of type
$$( (0,2)+(2,0))\otimes ((0,2)+(2,0)) + (2,2) = (4,0) + (2,2) + (0,4). $$
Now, we require $G$ to be an {\it integral } vector.
Define a $4\times 4$ matrix of integers so that
\eqn\fmint{
\eqalign{
G&= \pmatrix{w_1 & w_1^* & w_2 & w_2^*\cr} N \pmatrix{\tl w_1\cr \tl w_1^*\cr \tl w_2\cr \tl w_2^*\cr}\cr
& = N_{11} w_1\otimes \tl w_1+ N_{12}w_1\otimes \tl w_1^* + N_{13}w_1\otimes \tl w_2 + N_{14}w_1
\otimes \tl w_2^* + \cdots + N_{44} w_2^* \otimes \tl w_2^* \cr}
}
Now we have
\eqn\nfrxy{
 N = M^{tr}(a,a',b,c)\pmatrix{X & 0 \cr 0 & Y \cr}M(\tl a, \tl a',\tl b,\tl c)
}
where it is useful to define the matrix
\eqn\mmatrx{
M(a,a',b,c) = \pmatrix{1 & c/a' & 0 & 0 \cr 0 &-b/a'& a/a'& 1\cr
1 & - c/a' & b/a' & 0 \cr 0 & 0 & -a/a'& 1\cr}
}
so that
\eqn\trns{
\pmatrix{\nu_1 \cr \nu_2 \cr \gamma_1\cr \gamma_2 \cr} = M(a,a',b,c) \pmatrix{w_1 \cr
w_1^* \cr w_2 \cr w_2^*\cr}
}

Now we see that for any pair of attractor points in the complex structure
moduli space of $S\times \tilde S$, there are infinitely
many flux vacua leading to those specified points. To prove this
let us choose   $T,U\in Q(\sqrt{D})$
and $\tl T, \tl U \in Q(\sqrt{\tl D})$ to be concordant. Then   if $X,Y$ are
integer matrices divisible by $a'\tl a'$   the resulting matrix $N$ is a
matrix of integers. But, by construction, it leads to the specified
flux vacuum. For special values of $T,U,\tilde T, \tilde U$ in fact the
vacuum is not an isolated point. However, we expect that for generic
$T,U\in \IQ[\sqrt{D}]$  and $\tilde T, \tilde U \in \IQ[\sqrt{\tilde D}]$
the vacuum will be isolated. (We did not prove this rigorously.)

It should be stressed that there is no reason in the above construction for
the fields $\IQ[\sqrt{D}]$ and $\IQ[\sqrt{\tilde D}]$ to coincide. As we have
mentioned, by including further quantum corrections to the flux potential
one can associate an AdS vacuum to stationary points of \cnt\ with
 $G^{4,0} \not=0$. Moreover the scale
of the cosmological constant is, roughly speaking, governed by the value of the normalized
period \cnt\ with $\gamma = G$.
%
An easy computation shows that for the special vacua under consideration
\eqn\psecc{
\vert Z(G) \vert^2 = {a \tilde a \over a' \tilde a'}
\biggl \vert (x_{11} U - x_{21}) \tilde U - (x_{12} U - x_{22}) \biggr\vert^2
}
where $x_{ij}$ are the matrix elements of $X$ in \gmflx.
{}From this one learn that if one further imposes the condition that $G^{4,0}=0$ then, for generic
$X$, one finds that $U,\tl U$ must be in the same quadratic field. Moreover,
if $U$, $\tl U$ do {\it not} generate the same field then the distribution
of values of $\vert Z(G)\vert^2$, as $G$ runs over the different fluxes, is dense in $\IR$.

In physics, there is another constraint on the fluxes which severely cuts down
the above plethora of supersymmetric vacua. In the M-theoretic version the
net electric charge for the $C$-field must vanish on a compact space and therefore
\eqn\melt{
\int_{X_4} \half G^2 - {\chi(X_4)\over 24} + N_2 =0
}
where $N_2$ is the number of membranes, and, for supersymmetric vacua, is nonnegative.
Thus $[G]\cdot [G]$ is bounded. Equivalently, in  the IIB setup   the Bianchi identity on
the 5-form flux leads to a bound on   \KachruHE\
\eqn\neff{
N_f = \int_{X_3} F\wedge H  = {1\over \phi-\bar\phi} \int G_{IIB} \wedge G_{IIB}^*
}
As pointed out in   \AshokGK\ this leads to an important finiteness property:
{\it The number of flux vectors leading to vacua in a compact region of
moduli space is finite. }
Following \AshokGK\ let us prove this for the more general 4-fold problem. Let $\CK \subset
\CM_{cplx}(X_4)$ be a compact region in the moduli space of complex structures
of the elliptically fibered $X_4$. Consider $\CK \times H^4(X_4;\IR)$.
The subbundle of real vectors of type $H^{4,0} \oplus H^{2,2}_{primitive} \oplus H^{0,4} $
has a positive intersection product. With respect to a fixed basis on $H^4(X,\IR)$
the quadratic form is smoothly varying. Therefore,   the set of real vectors satisfying
$\half G^2\leq B$, for fixed bound $B$, is a compact set  in $\CK \times H^4(X_4;\IR)$ and hence projects
to a compact set $\CU$ in $ H^4(X_4;\IR)$. Therefore, there
can be at most a finite number of lattice vectors in  $\CU$. Note that it is
essential to use the primitivity condition.

As an example of how the bound $B$ imposes finiteness,
 consider the family \simplealso, with $F=n \gamma_1$,
$H= n \gamma_2$. Then $\int F\wedge H = 4n^2 b$. Since the denominators of $a,b$
must divide $n$, the denominators of $a,b$ are bounded when \neff\ is bounded.
Thus, the bound on \neff\ cuts down \simplealso\ to a finite set of examples. (In fact,
we may dispense with the cutoff $\CK$ on the   region in $\CM_{cplx}$.)

It would be interesting, even in the simple explicit examples above, to
give precise bounds for the number for flux vacua associated to a region $\CK$ and bound
$B$. This should be related to   class numbers. For example, the
solutions \simple\ have $N_f = 2 \vert D\vert$, and hence there are
$h(D)$ distinct such solutions. Unfortunately, the general relation
  appears to be complicated. For asymptotic estimates in the case of
general CY compactification,
under the assumption of uniform distribution, see \AshokGK\douglasdenef.

\subsec{ Conclusions}

Complex multiplication is  beautiful and profound.
Moreover, as we have shown,
arithmetic varieties related to number fields do seem
to be naturally selected in supersymmetric black holes,
F-theory, and flux compactifications.
The main open question, as far as the   author is
concerned, is whether the arithmetic of these varieties
has any important physical significance.


\bigskip
\noindent{\bf Acknowledgements:}
I would like to thank my collaborators on the work
which was reviewed above,  R. Dijkgraaf, J. Maldacena, S. Miller
A. Strominger, and E. Verlinde.
I also would like to
 thank F. Denef and E. Diaconescu for numerous detailed discussions
on the subject of lecture 2, and N. Yui for some helpful
correspondence. In addition I would like to thank B. Acharya,
A. Connes, F. Denef, R. Donagi, M. Douglas, S. Kachru, J. Lagarias, J. Markloff,
T. Pantev, K. Wendland,
and D. Zagier for useful comments and  discussions.
 I would also like to thank the Les Houches \'Ecole de Physique
for hospitality at the wonderful conference  and B. Julia and
P. van Hove  for the
invitation to speak at the conference. Finally, this work
 is supported in part by DOE grant DE-FG02-96ER40949.

\listrefs

\bye